%% file: SUS-15-006_temp.tex
\pdfoutput=1

\documentclass[11pt,twoside,a4paper,cmspaper,final,collab]{cms-tdr}
\def\svnVersion{385380}\def\cmsCernNoTag{CERN-EP/2016-239}\def\cmsCernDate{\today}\def\cmsMessage{Published in Physical Review D as \href{http://dx.doi.org/10.1103/PhysRevD.95.012011}{\doi{10.1103/PhysRevD.95.012011.}}}

\begin{document}\cmsNoteHeader{SUS-15-006}

\hyphenation{had-ron-i-za-tion}
\hyphenation{cal-or-i-me-ter}
\hyphenation{de-vices}
\RCS$Revision: 381691 $
\RCS$HeadURL: svn+ssh://svn.cern.ch/reps/tdr2/papers/SUS-15-006/trunk/SUS-15-006.tex $
\RCS$Id: SUS-15-006.tex 381691 2017-01-19 21:21:43Z alverson $
\newlength\cmsFigWidth
\ifthenelse{\boolean{cms@external}}{\setlength\cmsFigWidth{0.48\textwidth}}{\setlength\cmsFigWidth{0.75\textwidth}}
\ifthenelse{\boolean{cms@external}}{\providecommand{\cmsLeft}{top\xspace}}{\providecommand{\cmsLeft}{left\xspace}}
\ifthenelse{\boolean{cms@external}}{\providecommand{\cmsRight}{bottom\xspace}}{\providecommand{\cmsRight}{right\xspace}}
\ifthenelse{\boolean{cms@external}}{\providecommand{\CL}{C.L.\xspace}}{\providecommand{\CL}{CL\xspace}}
\ifthenelse{\boolean{cms@external}}{\providecommand{\NA}{\ensuremath{\cdots}\xspace}}{\providecommand{\NA}{\ensuremath{\text{---}}\xspace}}
\ifthenelse{\boolean{cms@external}}{\providecommand{\cmsTableResize[1]}{\relax{#1}}}{\providecommand{\cmsTableResize[1]}{\resizebox{\columnwidth}{!}{#1}}}
\ifthenelse{\boolean{cms@external}}{\providecommand{\ptvecell}{\ensuremath{\vec{p}_\mathrm{T}^{\kern2pt\ell}}\xspace}}{\providecommand{\ptvecell}{\ensuremath{\vec{p}_\mathrm{T}^{\ell}}\xspace}}
\newcommand{\LT}{\ensuremath{L_\mathrm{T}}\xspace}
\newcommand{\DF}{\ensuremath{\Delta \Phi}\xspace}
\newcommand{\Rcs}{\ensuremath{R_\mathrm{CS}}\xspace}
\newcommand{\Lp}{\ensuremath{L_\mathrm{P}}\xspace}
\newcommand{\nbjet}{\ensuremath{n_{\PQb}}\xspace}
\newcommand{\nbtag}{\ensuremath{n_{\PQb}}\xspace}
\newcommand{\njet}{\ensuremath{n_\text{jet}}\xspace}
\newcommand{\njetSR}{\ensuremath{n_\text{jet}^{\mathrm{SR}}}\xspace}
\newcommand{\kappab}{\ensuremath{\kappa_{\PQb}}\xspace}
\newcommand{\kappaW}{\ensuremath{\kappa_{\PW}}\xspace}
\newcommand{\kappatt}{\ensuremath{\kappa_{\ttbar}}\xspace}
\newcommand{\Wjets}{\ensuremath{\PW\text{+jets}}\xspace}
\newcommand{\Rcscorr}{\ensuremath{R_{\mathrm{CS}}^{\text{data(corr)}}}}
\newcommand{\ttjets}{\ensuremath{\ttbar\mathrm{+jets}}\xspace}

\cmsNoteHeader{SUS-15-006}
\title{Search for supersymmetry in events with one lepton and multiple jets in proton-proton collisions at \texorpdfstring{$\sqrt{s}=13$\TeV}{sqrt(s) = 13 TeV}}

\date{\today}

\abstract{
A search for supersymmetry is performed in events with a single electron or muon in
proton-proton collisions at a center-of-mass energy of 13\TeV.  The data were recorded
by the CMS experiment at the LHC and correspond to an integrated luminosity of 2.3\fbinv.
Several exclusive search regions are defined based on the number of jets and \PQb-tagged jets, the scalar sum of the jet transverse
momenta, and the scalar sum of the missing transverse momentum and the transverse momentum of the lepton.
The observed event yields in data are consistent with the expected backgrounds from standard model processes.
The results are interpreted using two simplified models of supersymmetric particle spectra, both of which describe gluino pair
production. In the first model, each gluino decays via a three-body process to top quarks and a neutralino, which is associated
with the observed missing transverse momentum in the event. Gluinos with masses up to 1.6\TeV are excluded for neutralino
masses below 600\GeV. In the second model, each gluino decays via a three-body process to two light quarks and a chargino,
which subsequently decays to a $\PW$ boson and a neutralino. The mass of the chargino is taken to be midway between the
gluino and neutralino masses. In this model, gluinos with masses below 1.4\TeV are excluded for
neutralino masses below 700\GeV.

}

\hypersetup{%
pdfauthor={CMS Collaboration},%
pdftitle={Search for supersymmetry in events with one lepton and multiple jets in proton-proton collisions at sqrt(s) = 13 TeV},%
pdfsubject={CMS},%
pdfkeywords={CMS, physics, supersymmetry}}

\maketitle

\section{Introduction}\label{sec:Introduction}
Supersymmetry (SUSY)~\cite{Ramond:1971gb,Golfand:1971iw,Neveu:1971rx,
Volkov:1972jx,Wess:1973kz,Wess:1974tw,Fayet:1974pd,Nilles:1983ge} is a well-motivated
theoretical framework that postulates new physics beyond the standard model (SM). Models
based on SUSY can address several open questions in particle physics, \eg the cancellation of quadratically divergent loop
corrections when calculating the squared mass of the Higgs boson. In $R$-parity~\cite{FARRAR1978575} conserving SUSY models, the
lightest SUSY particle (LSP) is stable and can be a viable dark matter candidate.
An inclusive search for SUSY in the single-lepton channel was performed with 13\TeV data
recorded in 2015 by the CMS experiment at the CERN LHC, corresponding to an integrated luminosity of 2.3\fbinv.
Similar searches were performed in 7\TeV~\cite{Chatrchyan:2012ola,Chatrchyan:2012sca,Aad:2012ms}
and in 8\TeV~\cite{Chatrchyan:2013iqa,Aad:2015mia,Aad:2014lra} data by the CMS and ATLAS experiments.
First results in the single-lepton final state at 13\TeV are also available from both collaborations~\cite{CMS-PAS-SUS-15-007,ATLAS-13TeV_single_lepton,ATLAS-13TeV_multib}.
In this paper, we present a search for gluino pair production
designed to be sensitive to a variety of SUSY models.

In this analysis, the main backgrounds arise from \Wjets events and top quark-antiquark (\ttjets) events, which also
lead to $\PW$-boson production. In \Wjets events, or in \ttjets events with a single leptonic $\PW$-boson decay, the missing
transverse momentum, \ptvecmiss, defined as the negative vector sum of the transverse momenta of all reconstructed
particles in the event, provides a measurement of the neutrino transverse momentum. The quantity
$\ptvecell + \ptvecmiss$, where $\ptvecell$ is the lepton transverse momentum vector, corresponds
to the transverse momentum of the $\PW$ boson in background events of this type. We also define the magnitude of
the missing transverse momentum, $\MET = |\ptvecmiss|$, and the sum  $\LT = \pt^{\ell} + \MET$, where $\pt^{\ell}$
is the magnitude of $\ptvecell$.

A key analysis variable is the azimuthal angle \DF, measured in the plane perpendicular to the beams, between
$\ptvecell$ and $\ptvecell + \ptvecmiss$.
In background events with a single $\PW$-boson decay, \DF corresponds to the azimuthal angle between the transverse
momentum vectors of the charged lepton and the $\PW$ boson. In such events, the distribution of \DF falls rapidly
and has a maximum value determined by the mass and transverse momentum of the $\PW$ boson. The higher the boost of the $\PW$ boson,
the smaller the maximum value of \DF. In SUSY events corresponding to our signal models, however, \MET typically
receives a large contribution from the missing momentum of the two neutralino LSPs. As a consequence, the \DF distribution
in signal events is roughly uniform. The main backgrounds can therefore be suppressed by rejecting events with a
small value of \DF. The primary remaining background arises from \ttjets production where both $\PW$ bosons decay into a
charged lepton and a neutrino, with one lepton being not well identified or falling outside the detector acceptance.
This background populates the high region of \DF.

Since many models of gluino pair production lead to final states with a large number of jets, the signal-to-background ratio
is very small in regions with low jet multiplicity. We therefore restrict the search to regions of large jet multiplicity
and use low jet multiplicity regions, dominantly populated by events from SM processes, to estimate the background.
Exclusive search regions are characterized
by the number of jets (\njet), the number of \PQb-tagged jets (\nbtag), the scalar
sum of the transverse momenta \pt of the jets (\HT), and \LT.

\begin{figure*}[tbp!]
  \includegraphics[width=0.35\textwidth]{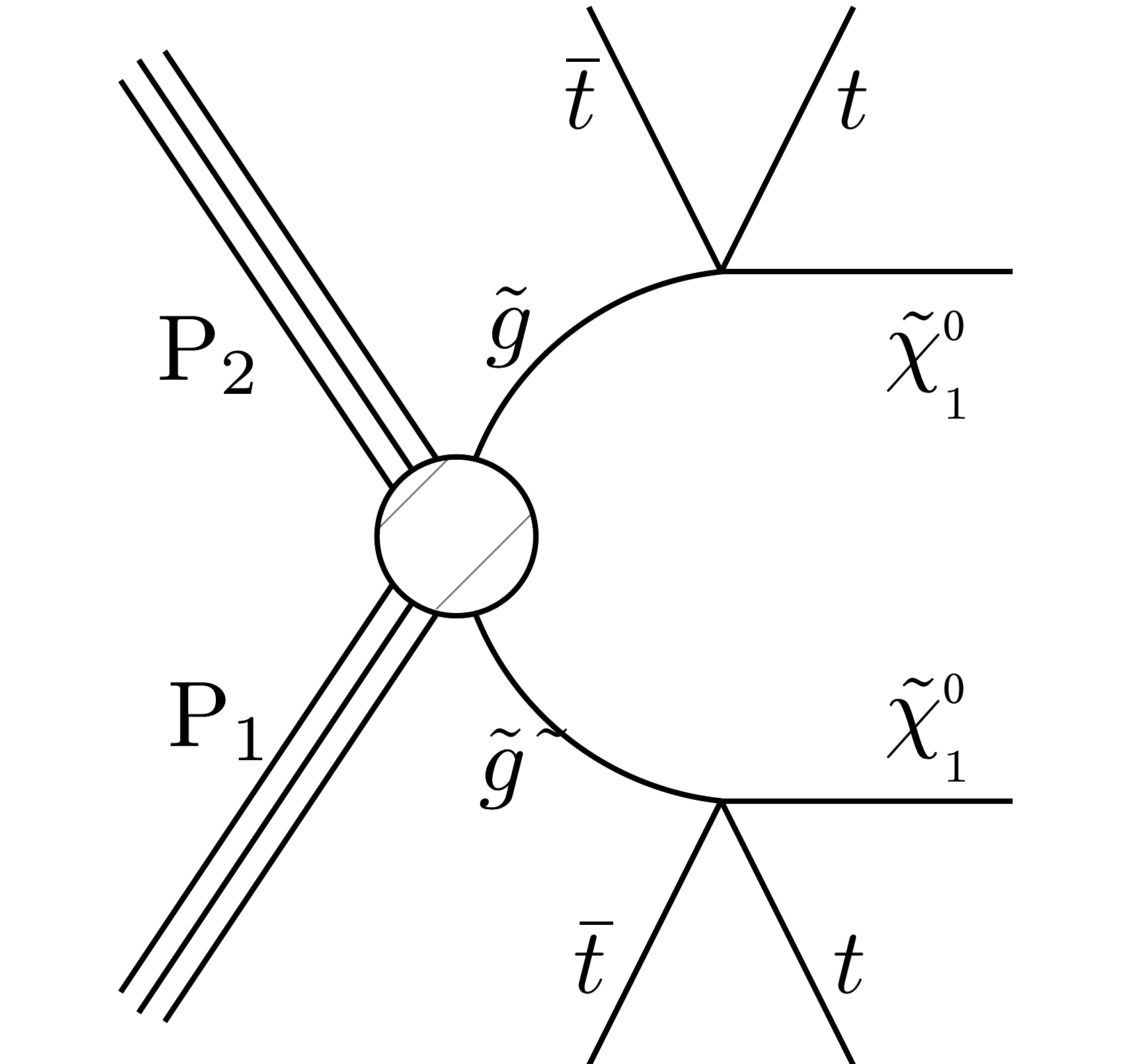}
  \hfil
  \includegraphics[width=0.35\textwidth]{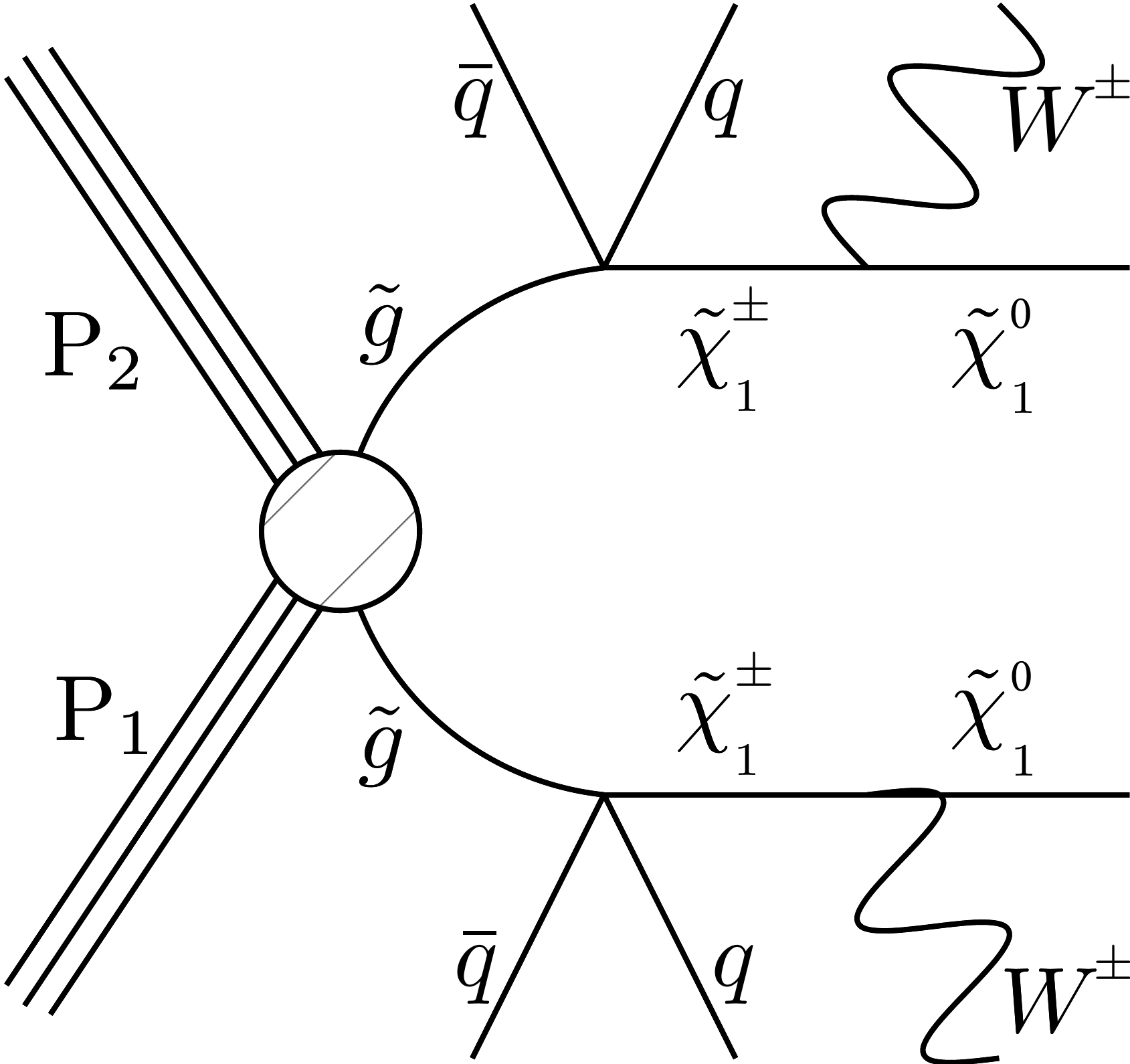}
  \hfil
\centering
  \caption{\label{fig:feynman_1} Diagrams showing the simplified models (left) T1tttt and
(right) T5qqqqWW. Depending on the mass difference between the
chargino (\PSGcpmDo) and the neutralino (\PSGczDo), the $\PW$ boson can be virtual.
  }
\end{figure*}

The results are interpreted in terms of simplified models~\cite{bib-sms-1,bib-sms-2,bib-sms-3,bib-sms-4} of gluino pair production.
In the first model, designated T1tttt and shown in Fig.~\ref{fig:feynman_1}~(left), gluinos are pair produced
and subsequently undergo three-body decays to $\ttbar + \PSGczDo$, where \PSGczDo is the lightest neutralino.
In the second model, termed T5qqqqWW and shown in Fig.~\ref{fig:feynman_1}~(right), the gluinos undergo three-body
decays to a quark-antiquark pair (\qqbar) from the first or second generation and a chargino (\PSGcpmDo). The
chargino mass is taken to be $m_{\PSGcpmDo}=0.5(m_{\PSg}+m_{\PSGczDo})$. The chargino then decays to a $\PW$ boson
and the \PSGczDo, where the $\PW$ boson can be virtual, depending on the mass difference between the chargino and
the lightest neutralino.

The organization of this paper is as follows.
Section~\ref{sec:detector} describes the CMS detector.
The event reconstruction and selection are discussed in Sections~\ref{sec:reco} and \ref{sec:eventselection},
respectively. The background estimations are given in Section~\ref{sec:background_estimation}. An overview of
the main systematic uncertainties is presented in Section~\ref{sec:systematics}.
The results are discussed and interpreted in Section~\ref{sec:results_interpretation},
and a summary is given in Section~\ref{sec:summary}.

\section{The CMS detector}\label{sec:detector}

The central feature of the CMS apparatus is a superconducting solenoid of 6\unit{m} internal
diameter, providing a magnetic field of 3.8\unit{T}. A silicon pixel and strip tracker, a lead
tungstate crystal electromagnetic calorimeter (ECAL), and a brass and scintillator hadron calorimeter (HCAL), each composed of a barrel and two endcap sections, reside within the solenoid volume. Forward calorimeters extend the
pseudorapidity ($\eta$)~\cite{Chatrchyan:2008zzk} coverage provided
by the barrel and endcap detectors.
Muons are measured in the range $\abs{\eta}< 2.4$, with detection planes made using three
technologies: drift tubes, cathode strip chambers, and resistive plate chambers.

The silicon tracker measures charged particles within the range $\abs{\eta}< 2.5$. Isolated
particles with transverse momenta $\pt = 100\GeV$, emitted at $\abs{\eta} < 1.4$, have track resolutions of 2.8\% in
\pt, and 10 (30)\mum in the transverse (longitudinal) impact parameter~\cite{TRK-11-001}. The ECAL and HCAL measure
energy depositions in the range $\abs{\eta} < 3$, with quartz fibre and steel forward calorimeters extending the
coverage to $\abs{\eta} < 5$. When information from the various detector systems is combined, the resulting jet energy
resolution is typically 15\% at 10\GeV, 8\% at 100\GeV, and 4\% at 1\TeV~\cite{Chatrchyan:2011ds}. The momentum resolution for
electrons with $\pt \approx 45$\GeV from $\Z \to \Pe \Pe$ decays ranges from 1.7\% for
electrons that do not shower in the barrel region, to 4.5\% for electrons that shower in the
endcaps~\cite{Khachatryan:2015hwa}. Matching muons to tracks
measured in the silicon tracker yields relative transverse momentum resolutions for muons with
$20 <\pt < 100\GeV$ of 1.3--2.0\% in the barrel, and less than 6\% in the endcaps. The \pt resolution in the
barrel is below 10\% for muons with \pt up to 1\TeV~\cite{Chatrchyan:2012xi}.

The CMS trigger system consists of two levels, where the first level (L1), composed of custom hardware
processors, uses information from the calorimeters and muon detectors to select the most interesting
events in a fixed time interval of less than 4\mus. The high-level trigger (HLT) processor farm further
decreases the event rate from around 100\unit{kHz} to less than 1\unit{kHz}, before data storage.

A more detailed description of the CMS detector, together with a definition of the coordinate system
used and the relevant kinematic variables, can be found in Ref.~\cite{Chatrchyan:2008zzk}.
\section{Event reconstruction and simulation} \label{sec:reco}

All objects in the event are reconstructed using the particle-flow event reconstruction
algorithm~\cite{CMS-PAS-PFT-09-001,CMS-PAS-PFT-10-001}, which reconstructs and identifies
each individual particle through an optimized combination of information from the various
elements of the CMS detector.
The energy of electrons is determined from a combination of the electron momentum at the primary interaction vertex
as determined by the tracker, the energy of the corresponding ECAL cluster, and the energy sum of all bremsstrahlung
photons spatially compatible with originating from the electron track~\cite{Khachatryan:2015hwa}. Electron candidates
are required to satisfy identification criteria designed to suppress contributions from misidentified jets, photon
conversions, and electrons from heavy-flavor quark decays.
Muons are reconstructed using a stand-alone muon track in the muon system serving as a seed to find a corresponding
track in the silicon detector~\cite{Chatrchyan:2012xi}. Additional criteria include requirements on the track and hit parameters.
Events are vetoed if additional electrons or muons with looser identification requirements are found.

The degree of isolation of a lepton from other particles provides a strong indication of whether it was produced
in a hadronic jet, such as a jet resulting from the fragmentation of a \PQb quark, or in the leptonic decay of a $\PW$
boson or other heavy particle. Lepton isolation is quantified by performing a scalar sum of the transverse momenta
of all particles that lie within a cone of specified size around the lepton momentum vector, excluding the contribution
of the lepton itself. To maintain high efficiency for signal events, which typically contain a large number of
jets from the SUSY decay chains, we use a \pt-dependent cone radius
$R=(0.2,\,10\GeV / \pt[\GeVns{}],\,0.05)$ for $(\pt < 50\GeV,\,50\GeV < \pt < 200\GeV,\,\pt > 200\GeV)$, respectively.
The isolation variable is defined as a relative quantity, $I_\text{rel}$,  by dividing this scalar sum by the \pt of the lepton. For selected muons or electrons, we require $I_\text{rel}<0.2$ and $I_\text{rel}<0.1$,
respectively, while for additional leptons used in the event veto, we require $I_\text{rel}<0.4$. When computing the
isolation variable, an area-based correction is applied to remove the contribution of particles from additional proton-proton
collisions within the same or neighboring bunch crossings (pileup).

The energy of charged hadrons is determined from a combination of their
momenta measured in the tracker and the matching ECAL and HCAL energy depositions, corrected for zero-suppression
effects in the readout electronics, and for the response function of the calorimeters to hadronic showers. Finally, the energy of neutral hadrons is obtained from the corresponding corrected ECAL and HCAL energies.

Jets are clustered with the anti-\kt algorithm~\cite{Cacciari:2008gp} with a distance parameter
of 0.4~\cite{Chatrchyan:2011ds}, as implemented in the \FASTJET package~\cite{Cacciari:2011ma}. Jet momentum is determined as
the vectorial sum of all particle momenta in the jet. An offset is subtracted from the jet energies
to take into account the contribution from pileup~\cite{Cacciari:2007fd}. Jet energy corrections are
obtained from simulation, and are confirmed with in situ
measurements of the energy balance in dijet and photon+jet events~\cite{Chatrchyan:2011ds}. Additional selection criteria
are applied to each event to remove spurious jet-like features originating from isolated noise patterns in certain HCAL
regions.

To identify jets originating from \PQb quarks, we use an inclusive combined secondary vertex tagger (CSVv2)~\cite{Chatrchyan:2012jua,CMS-PAS-BTV-15-001},
which employs both secondary vertex and track-based information. The working point is chosen to have about 70\% \PQb tagging
efficiency and a $1.5$\% light-flavor misidentification rate~\cite{CMS-PAS-BTV-13-001}.
Double counting of objects is avoided by not considering jets that lie within a cone of radius
0.4 around a selected lepton.

While the main backgrounds are determined from data, as described in Section~\ref{sec:background_estimation}, simulated
events are used to validate the techniques, and to estimate extrapolation factors as needed. In addition,
some smaller backgrounds are estimated entirely from simulation. The leading-order (LO)
{\MADGRAPH}5~\cite{Alwall:2011uj} event generator, using the NNPDF3.0LO~\cite{Ball:2014uwa}
parton distribution functions (PDFs), is used to simulate \ttjets, \Wjets, Z+jets, and multijet events.
Single-top quark events in the $t$-channel and the $\cPqt\PW$ process are generated using the next-to-leading order (NLO) {\POWHEG}v1.0~\cite{Nason:2004rx,Frixione:2007vw,Alioli:2010xd,Alioli:2009je,Re:2010bp} program,
and in the $s$-channel process, as well as for $\ttbar$W and $\ttbar$Z production, using NLO {\MADGRAPH}5{\textunderscore}a{\MCATNLO}~\cite{Alwall:2014hca}.
All signal events are generated with {\MADGRAPH}5, with up to two partons in addition to the gluino pair.
Both programs use the NNPDF3.0NLO~\cite{Ball:2014uwa} PDF.
The gluino decays are based on a pure phase-space matrix element~\cite{Sjostrand:2014zea}, with signal production cross
sections~\cite{bib-nlo-nll-01,bib-nlo-nll-02,bib-nlo-nll-03,bib-nlo-nll-04,bib-nlo-nll-05}
computed at NLO plus next-to-leading-logarithm (NLL) accuracy.

We define several benchmark points:
the model T1tttt(1.2,0.8) (T1tttt(1.5,0.1)) corresponds to a gluino mass of
1.2 (1.5)\TeV and neutralino mass of 0.8 (0.1)\TeV, respectively. The model T5qqqqWW(1.0,0.7)
(T5qqqqWW(1.2,0.8) and T5qqqqWW(1.5,0.1)) corresponds to a gluino mass of
1.0 (1.2 and 1.5)\TeV and neutralino mass of 0.7 (0.8 and 0.1)\TeV. For the latter, the intermediate
chargino mass is fixed at 0.85 (1.0 and 0.8)\TeV.

Showering and hadronization of all partons is performed using the \PYTHIA~8.2 package~\cite{Sjostrand:2014zea}.
Pileup is generated for some nominal distribution of the number of proton-proton interactions per bunch crossing,
which is weighted to match the corresponding distribution in data.
The detector response for all backgrounds is modelled using the \GEANTfour~\cite{Agostinelli:2002hh} package,
while for the signal the CMS fast simulation program~\cite{bib-cms-fastsim-02} is used to reduce
computation time.
The fast simulation has been validated against the detailed \GEANTfour-based simulation for the variables relevant for this
search, and efficiency corrections based on measurements in data are applied.

\section{Trigger and event selection}\label{sec:eventselection}

The events are selected with an L1 trigger requiring $\HT>150\GeV$, followed by HLT requirements of $\HT>350\GeV$
(online reconstruction) and at least one isolated lepton (an electron or muon) satisfying $\pt>15\GeV$.
A trigger efficiency of 94$\pm$1\% is observed in the kinematic regime of the analysis, defined by lepton $\pt>25\GeV$
and $\HT>500\GeV$, where the trigger efficiency reaches its maximum.

The electron or muon candidate is required to have a minimum \pt of 25\GeV. Events with additional electrons
or muons with $\pt > 10\GeV$, satisfying the criteria for vetoed leptons, are rejected.
Jets are selected with $\pt > 30\GeV$ and $|\eta| < 2.4$.
In all search regions we require at least five jets, where the two highest-\pt jets must satisfy $\pt > 80\GeV$.

To separate possible new-physics signals from background, we use the \LT variable, which is
defined as the scalar sum of the lepton \pt and the missing transverse energy \MET, and reflects the
leptonic energy scale of the event. A minimum \LT of 250\GeV is required, such that the analysis is
not only sensitive to events with high \MET, but also to signal events with very small \MET, but higher
lepton \pt. An additional kinematic quantity important for the search is given by the hadronic energy scale
of the event \HT. A cutflow for the benchmark signal models is given in Table~\ref{tab:baselineEventYields}.

\begin{table*}
\centering
\topcaption{Expected event yields for SUSY signal benchmark models, normalized to
2.3\fbinv. The baseline selection corresponds to all requirements up to and including the requirement on \LT. The last
two lines are exclusive for the zero-\PQb and the multi-\PQb selection, respectively. The events are corrected with scale factors
to account for differences in the lepton identification and isolation
efficiencies, trigger efficiency, and the \PQb-tagging efficiency between simulation and data.
\label{tab:baselineEventYields} }

\begin{scotch}{ l c c c c }
\multirow{2}{*}{Selection}    & T1tttt  & T1tttt   & T5qqqqWW    & T5qqqqWW  \\
     &  (1.2,0.8) &  (1.5,0.1)  &  (1.2,0.8)   &  (1.5,0.1) \\

\hline

All events             &   178   &   30  & 185 & 31 \\
One hard lepton           &   55    &   11      & 51& 9.3 \\
No veto lepton         &   45    &   9.1     & 47& 8.8 \\
$\njet \geq 5$              &   44    &   8.9     & 36& 8.1 \\
$p_{T}(\text{jet 2}) > 80\GeV$         &   36    &   8.9     &34 & 8.1 \\
$\HT > 500\GeV$            &   30    &   8.9     & 27& 8.1 \\
$\LT > 250\GeV$            &   15    &   8.4     & 21 &  7.8 \\  \hline
$\nbjet=0$ and $\DF > 0.75$    &   0.47    &   0.26   & 11 & 3.5\\  \hline
$\nbjet\ge1$, $\njet\geq6$  and $\DF > 0.75$     &    9.3    &   5.1    & 2.9 & 1.2 \\
\end{scotch}
\end{table*}

After imposing the minimum requirements on \LT and \HT, several search regions are defined in bins of
\njet, \nbtag, \LT, and \HT, where \njet and \nbtag are the numbers of jets and \PQb-tagged jets, respectively.
Defining search bins in \PQb-jet multiplicity enables the analysis to target specific event
topologies and to separate them from SM backgrounds. The phase space is divided into exclusive
[0, 1, 2, $\geq$3] \PQb-tagged jet categories when defining search bins, with a minimum \PQb-jet $\pt$ of $30\GeV$.

\begin{figure}[tbp!]
\centering
\includegraphics[width=0.45\textwidth]{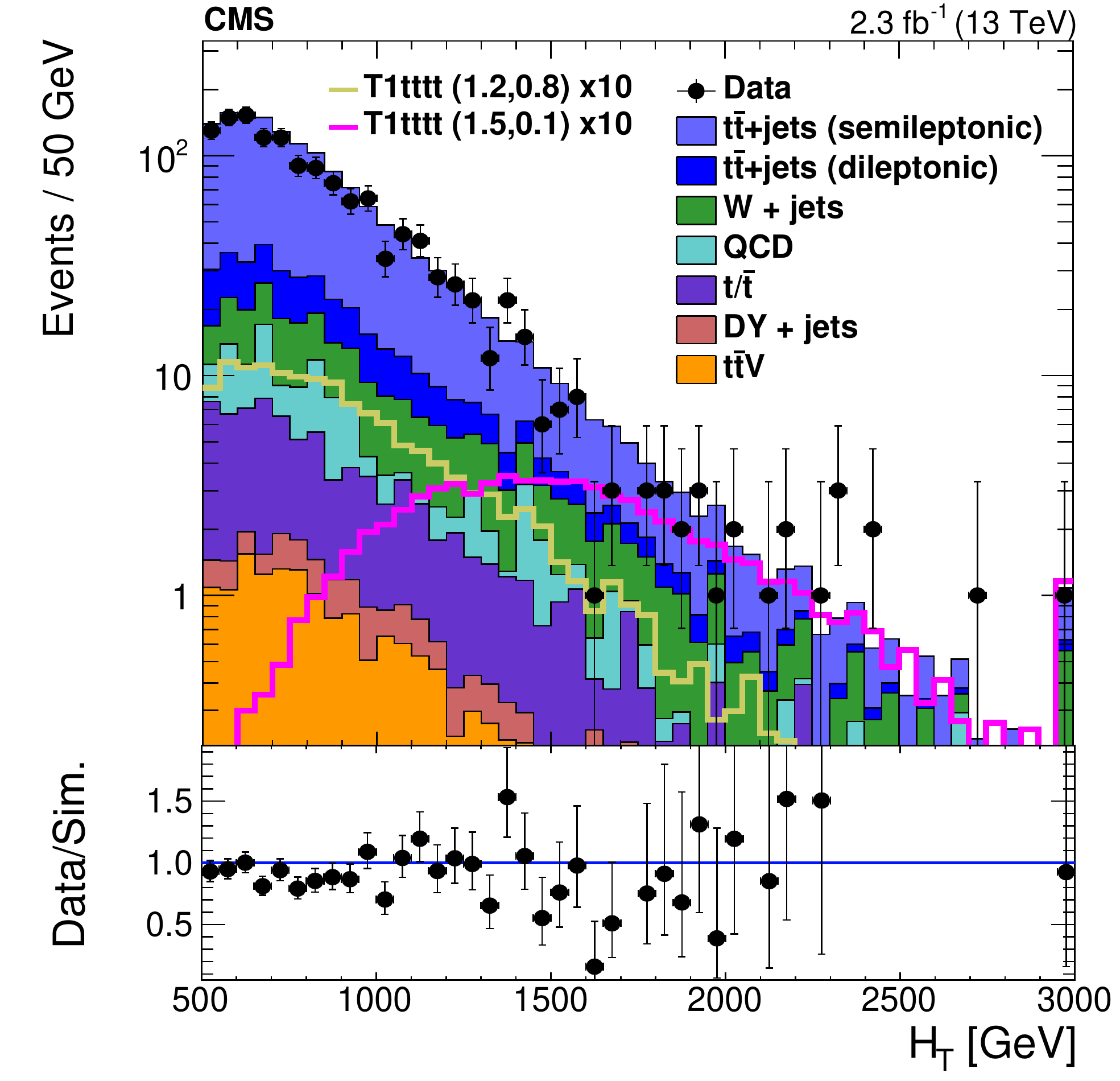}
\includegraphics[width=0.45\textwidth]{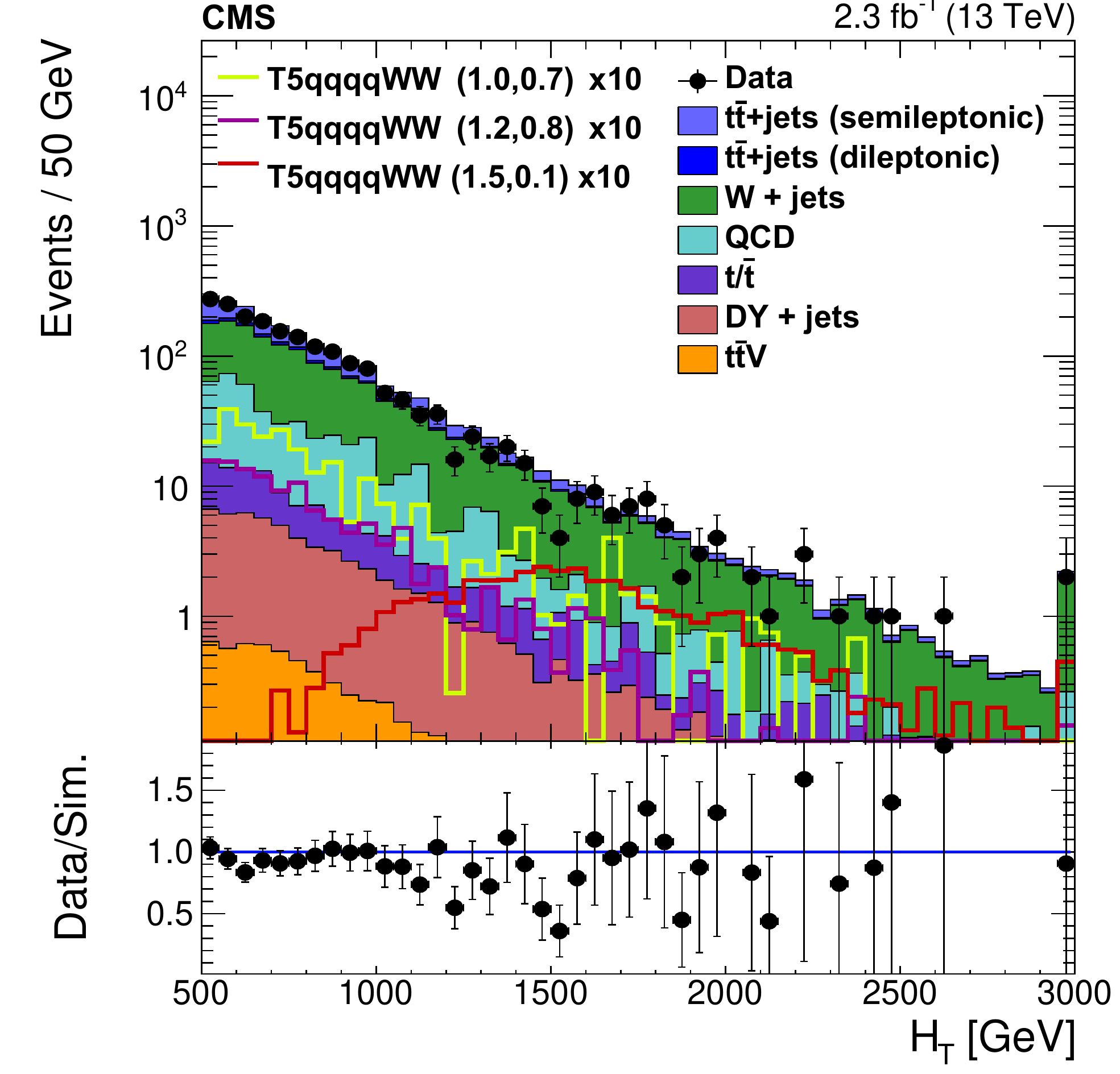}
\caption{The \HT distribution for (\cmsLeft) the multi-\PQb analysis and (\cmsRight) the zero-\PQb analysis, both after the baseline selection.
The simulated background events are stacked on top of each other, and several signal points are overlaid for illustration,
but without stacking. Overflows are included in the last bin. The label DY refers to $\qqbar\to \Z/\gamma^* \to
\ell^{+}\ell^{-}$ events, and QCD refers to multijet events. The event
yields for the benchmark models have been scaled up by a factor of 10.
The ratio of data to simulation is given below each of the panels.
All uncertainties are statistical only.
}
\label{fig:HT_LT}
\end{figure}

All search bins with at least one \PQb-tagged jet, called in the following ``multi-\PQb'' bins, are sensitive to the T1tttt model,
while the search bins requiring zero \PQb-tagged jets, called ``zero-\PQb'' bins, are sensitive to the T5qqqqWW model. The
baseline selection and the background estimation method differ for these two \PQb-tag categories.
For T1tttt, we expect a large number of jets and find in simulation that the \njet distribution
peaks at eight jets for most mass points. We require at least six jets for the multi-\PQb analysis and define two independent
categories with 6--8 and $\geq$9 jets. For the zero-\PQb analysis, where the investigated simplified T5qqqqWW model
has fewer jets, we require in the search region 5, 6--7, or $\geq$8 jets.
Depending on the specific SUSY particle masses, the hadronic event activity varies. To accommodate this, we define search bins in
\HT. Figure~\ref{fig:HT_LT} shows the \HT distributions for the multi-\PQb and the zero-\PQb selection. To exploit the strong separation
power associated with the \LT variable, we divide the search region into four bins in \LT, such that sufficient statistical
accuracy is given in each control bin to predict the background in the corresponding search bin.

\begin{figure}[tbp!]
\centering
\includegraphics[width=0.48\textwidth]{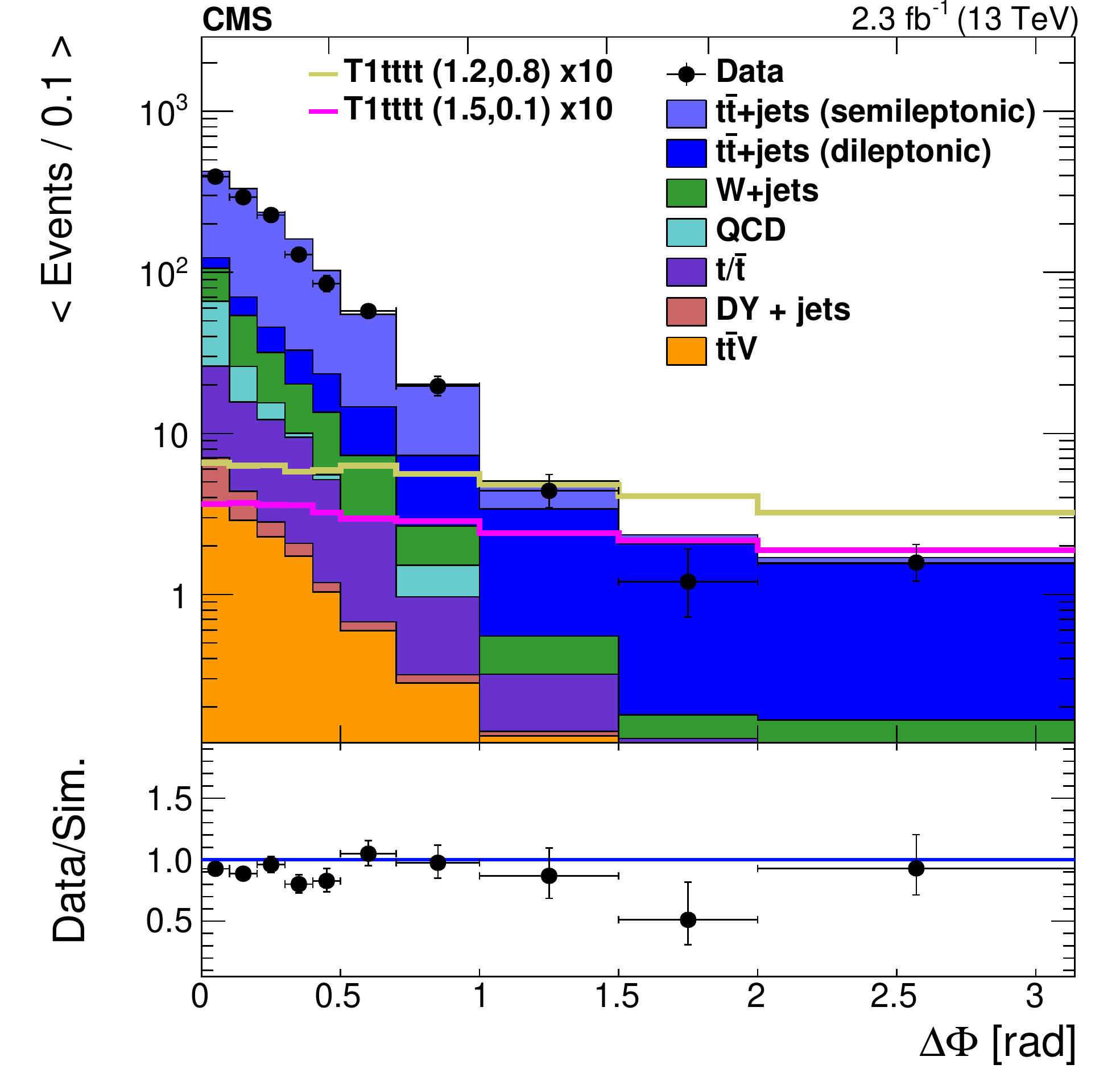}
\includegraphics[width=0.48\textwidth]{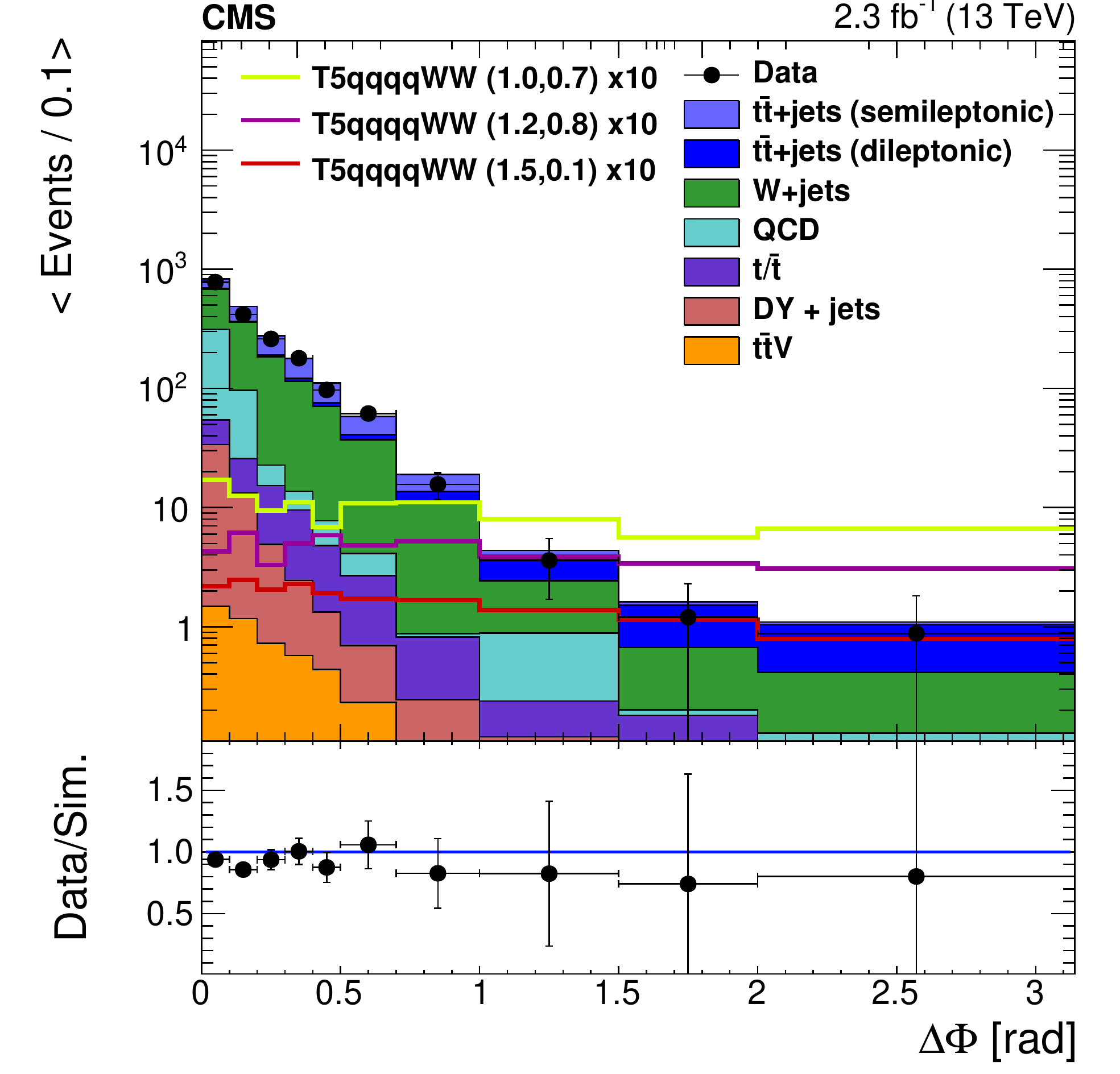}
\caption{Comparison of the \DF distribution for (\cmsLeft) the multi-\PQb and (\cmsRight) the zero-\PQb analysis after the baseline selection.
The simulated background events are stacked on top of each other, and several signal points are overlaid for illustration,
but without stacking. The wider bins are normalized to a bin width of 0.1. The label DY refers to $\qqbar \to \Z/\gamma^* \to
\ell^{+}\ell^{-}$ events, and QCD refers to multijet events. The event yields for the benchmark models have been scaled
up by a factor of 10.
The ratio of data to simulation is given below each of the panels.
}
\label{fig:dPhi}
\end{figure}

After these selections, the main backgrounds are leptonically decaying \Wjets and semi-leptonic \ttbar events.
These backgrounds, both of which contain one lepton and one neutrino (from the $\PW$ boson decay) in the final state,
are mostly located at small \DF values due to the correlation between the lepton and the
neutrino. Therefore, the region with large \DF is defined as the search region,
while the events with small \DF are used as the control sample. Figure~\ref{fig:dPhi} shows the \DF distributions
for the zero-\PQb and multi-\PQb search regions.
The ratio of the background event yield in the search region to that in the control region is determined
in the corresponding signal-depleted sideband regions, which have smaller values of \njet, as discussed in
Section~\ref{sec:background_estimation}.
Since the angle between the $\PW$ boson and the lepton depends on the W momentum, being smaller for $\PW$ bosons with higher boost, the
\DF requirement for the signal region is chosen depending on \LT, which is a measure of the $\PW$ boson \pt.
For the zero-\PQb analysis, \DF is required to be
larger than 1.0 for most regions except for those with large \LT, where the requirement is relaxed to 0.75, while the multi-\PQb
analysis has a relaxed \DF requirement of 0.75 and 0.5 for medium- and high-\LT regions, respectively.

In total, we define 30 search bins in the multi-\PQb analysis and 13 search bins in
the zero-\PQb analysis, as described in detail in Table~\ref{tab:sigreg}.

\begin{table*}[!htb]
\topcaption{Search regions and the corresponding minimum \DF requirements.}
\label{tab:sigreg}
\centering
\begin{scotch}{ccccc}
\njet &\nbtag & \LT$[\GeVns{}]$ & \HT$[\GeVns{}]$ & \DF[rad] \\\hline
[6,8]& \multirow{2}{*}{${=}1$, ${=}2$, $\geq$3} & $[250,350]$ & $[500,750]$, $\geq$750&$1.0$\\
& &$[350,450]$&$[500,750]$, $\geq$750&\multirow{2}{*}{$0.75$}\\
& \multirow{2}{*}{${=}1$, $\geq$2} &$[450,600]$&$[500,1250]$, $\geq$1250&\\
& &$\geq$600& $[500,1250]$, $\geq$1250&$0.5$\\[2ex]

$\geq$9 & ${=}1$, ${=}2$ &$[250,350]$ & $[500,1250]$, $\geq$1250 & \multirow{2}{*}{$1.0$} \\
 & $\geq$3& & $\geq$500 & \\
& ${=}1$, ${=}2$, $\geq$3 & $[350,450]$ &$\geq$500&\multirow{2}{*}{$0.75$} \\
& ${=}1$, $\geq$2 & $\geq$450 &$\geq$500& \\[2ex]

5 & \multirow{5}{*}{0} &  $[250,350]$, $[350,450]$, ${\geq}$450 & ${\geq}$500 & \multirow{2}{*}{ 1.0}\\[2ex]

[6$,$7]& & $[250,350]$, $[350,450]$ & $[500,750]$, ${\geq}$750 & \\
& & ${\geq}$450 & $[500,1000]$, ${\geq}$1000 & 0.75\\[2ex]

$\geq$8& & $[250,350]$  & $[500,750]$, ${\geq}$750 & 1.0\\
 & & $[350,450]$, ${\geq}$450  & ${\geq}$500 & 0.75\\
\end{scotch}
\end{table*}

\section{Background estimation}\label{sec:background_estimation}

The dominant backgrounds in this search are from \ttjets and \Wjets events, whose contributions vary with the multiplicity
of \PQb-tagged jets and the kinematic region in \HT and \LT.
To determine these backgrounds, we define two regions for each bin in \LT, \HT, and \nbtag: the search region (SR)
with large values of \DF, and the control region (CR) with low values of \DF, with the separation
requirement depending on the \LT value, as shown in Table~\ref{tab:sigreg}.
We further divide each of these bins into low-\njet sideband (SB) and high-\njet
main band (MB) regions.

About 10--15\% of the SM background events in the CR are expected to be multijet events (denoted in the following
as QCD), and are predicted as described in Section~\ref{sec:qcdest}. Since the multijet background is negligible in
the SR, it is subtracted from the number of background events in the CR when calculating the transfer factor
$\Rcs^\text{data}$ to extrapolate from CR (low-\DF) to SR (high-\DF). This transfer factor $\Rcs^\text{data}$ is determined
from data in the low-\njet SB regions, separately for each \LT, \HT, and \nbtag search region:
\begin{equation}
\Rcs^\text{data} = \frac{N_\text{data}^\mathrm{SB}(\mathrm{SR})}{N_\text{data}^\mathrm{SB}(\mathrm{CR}) - N_{\mathrm{QCD}\; \text{pred}}^{\mathrm{SB}}(\mathrm{CR})},
\label{eq:Rcs}
\end{equation}
where $N_\text{data}^\mathrm{SB }(\mathrm{SR})$ is the number of events in the low-\njet SB high-\DF signal region, $N_\text{data}^\mathrm{SB}(\mathrm{CR})$ the number of events
in the low-\njet SB low-\DF control region, and  $N_{\mathrm{QCD}\; \text{pred}}^\mathrm{SB}(\mathrm{CR})$ the predicted number of QCD multijet events in the SB CR.

In the regions with one \PQb tag and four or five jets, about 80\% \ttjets events and 15--20\% \Wjets and single
top quark events are expected, while in all other multi-\PQb regions, \ttbar background is completely dominant.
Because only a single SM background dominates in the multi-\PQb analysis, just one \Rcs factor is needed for each \LT, \HT,
and \nbtag range. In the zero-\PQb bins, the contributions from \Wjets and \ttjets are roughly equal.
Here, an extension of the multi-\PQb strategy is employed, which takes into account differences in the \Rcs values
for these two backgrounds.

An overview of the $(\njet, \nbjet)$ regions used in this analysis, as
discussed in detail in the following Sections~\ref{sec:rcs_multib} to \ref{sec:qcdest}, is given in Table~\ref{tab:Regions}.

\begin{table*}[!htb]
\topcaption{Overview of the definitions of sideband and mainband regions. For the multijet (QCD) fit the electron (e) sample is used,
while for the determination (det.) of $\Rcs(\Wpm)$ the muon ($\mu$) sample is used. Empty cells are not used in this analysis.
  }
\centering
\begin{scotch}{ccc{c}@{\hspace*{5pt}}cc}
Analysis       & \multicolumn{2}{c}{Multi-\PQb analysis} && \multicolumn{2}{c}{Zero-\PQb analysis} \\ \cline{1-3}\cline{5-6}
               & $\nbtag = 0$ & $\nbtag \geq 1$ && $\nbtag = 0$                & $\nbtag = 1$ \\
$\njet = 3$    &   QCD bkg.\ fit (\Pe\ sample) &&& $\Rcs(\Wpm)$ det. ($\mu$ sample), &        \\
$\njet = 4$    &  QCD bkg.\ fit (\Pe\ sample) &\Rcs det.&&  QCD bkg.\ fit (\Pe\ sample) & $\Rcs(\ttjets)$ det. \\
$\njet = 5$    & & \Rcs det.&& MB & $\Rcs(\ttjets)$ det. \\ 
$\njet \geq 6$ & & MB &&MB & \\
\end{scotch}
\label{tab:Regions}
\end{table*}

\subsection{Estimate of the leading backgrounds for \texorpdfstring{$\nbtag \ge 1$}{no. b > 1}} \label{sec:rcs_multib}

For the multi-\PQb analysis, the SB region, where \Rcs is determined, is required to have four or five jets, while the
MB region must satisfy $\njet \in [6-8]$ or $\njet \geq 9$. To account for possible
differences in this extrapolation from SB to MB as a function of jet multiplicity, we apply multiplicative
correction factors $\kappa_\mathrm{EW}$, determined from simulation.
The predicted number $N^\mathrm{MB}_\text{pred} (\mathrm{SR})$ of background events in each MB SR is then given by:
\begin{equation}
N^\mathrm{MB}_\text{pred}(\mathrm{SR}) = \Rcs^\text{data} \, \kappa_\mathrm{EW} \, \left[ N_\text{data}^\mathrm{MB}(\mathrm{CR}) - N^\mathrm{MB}_{\mathrm{QCD}\; \text{pred}}(\mathrm{CR}) \right],
\label{eq:prediction}
\end{equation}
with
\begin{equation}
\kappa_\mathrm{EW} = \frac{\Rcs^\mathrm{MC}(\mathrm{MB, EW})}{\Rcs^\mathrm{MC}({\mathrm{SB}, \mathrm{EW}})}.
\label{eq:kappaewk}
\end{equation}
Here $\Rcs^\text{data}$ is determined from Eq.~(\ref{eq:Rcs}), $N_\text{data}^\mathrm{MB}(\mathrm{CR})$ is the number of data events in the
CR of the MB region, and $N^\mathrm{MB}_{\mathrm{QCD}\; \text{pred}}(\mathrm{CR})$ is the predicted number of multijet events in the MB. The label EW refers
to all backgrounds other than multijets.
The residual difference of the values of \Rcs between the SB and MB regions is evaluated in simulation as the correction
factor $\kappa_\mathrm{EW}$ given by Eq.~(\ref{eq:kappaewk}), where $\Rcs^\mathrm{MC}({\mathrm{MB}, \mathrm{EW}})$ is the \Rcs in
a search MB region from simulation and $\Rcs^\mathrm{MC}({\mathrm{SB}, \mathrm{EW}})$ is the \Rcs in the corresponding SB region
in simulation for the EW background.

The $\kappa_\mathrm{EW}$ factor is determined separately for each search bin, except that an overall $\kappa_\mathrm{EW}$-factor
is applied for the $\nbtag \ge 2$ search bins with the same \HT and \LT, since
the $\kappa_\mathrm{EW}$ factors are found to be nearly independent of \nbtag.
Similarly, \Rcs at very high \HT is determined jointly across all three \nbtag bins to increase the number of events,
as the overall uncertainty of the background prediction for several of the search bins is dominated by the statistical
uncertainty of the yield in the CR of the main band.

The value of \Rcs for the total background is equal to the sum of the \Rcs values of each background component,
weighted with the relative contributions of the components.
For semileptonic \ttbar and \Wjets events, which contain both one neutrino from the hard interaction, \Rcs typically has values of
0.01 to 0.04, depending on the search bin.
In events with more than one neutrino, \eg in \ttbar events in which both $\PW$ bosons decay leptonically, \Rcs is higher
with values of around 0.5. This is visible in Fig.~\ref{fig:dPhi}, where at high \DF
a large fraction of events is due to dileptonic \ttjets background, while the low-\DF region is dominated by events
with only one neutrino. A larger \Rcs is also expected for events with three neutrinos, such as $\ttbar$Z, when the
\ttbar system decays semileptonically and the \Z boson decays to two neutrinos. The influence of these latter processes is small, since
their relative contribution to the background is minor.
Most of the SRs with six or more jets are dominated by semileptonic \ttbar events, and therefore this background
dominates the total \Rcs value of $\approx$0.05. As the \Rcs for dileptonic \ttbar events is an order of magnitude larger than for
semileptonic \ttbar events, a slight change in composition in the CR from low- to high-\njet multiplicity translates
into $\kappa_\mathrm{EW}$ slightly different from unity. This change in the dileptonic \ttbar contribution is accounted for by assigning
an uncertainty on the \njet extrapolation based on a dileptonic control sample in data, as discussed in Section~\ref{sec:systematics}.

\subsection{Estimate of the leading backgrounds for \texorpdfstring{$\nbtag = 0$}{no. b=0}} \label{sec:rcs_0b}

For search bins in which \PQb-tagged jets are vetoed, the background contributions from \Wjets and \ttjets events are estimated
by applying the \Rcs method separately to each of the two components.
This strategy implies the use of two sidebands enriched in \Wjets and \ttjets events, respectively.
We write the total background in each search region \njetSR (with a \DF  requirement as shown in Table~\ref{tab:sigreg}) as:
\begin{equation}
N_\mathrm{MB}^\mathrm{SR}(0\PQb) = N^\mathrm{SR}_{\PW}(0\PQb)+N^\mathrm{SR}_{\ttbar}(0\PQb) +N^\mathrm{SR(MC)}_{\text{other}}(0\PQb),
\end{equation}
where the predicted yields of \Wjets and \ttjets background events are denoted by  $N^\mathrm{SR}_{\PW}$ and $N^\mathrm{SR}_{\ttbar}$, respectively.
Additional backgrounds from rare sources are estimated from simulation and denoted by $N^\mathrm{SR(MC)}_{\text{other}}$.

\begin{figure}[tbp!]
\centering
    \includegraphics[width=0.48\textwidth]{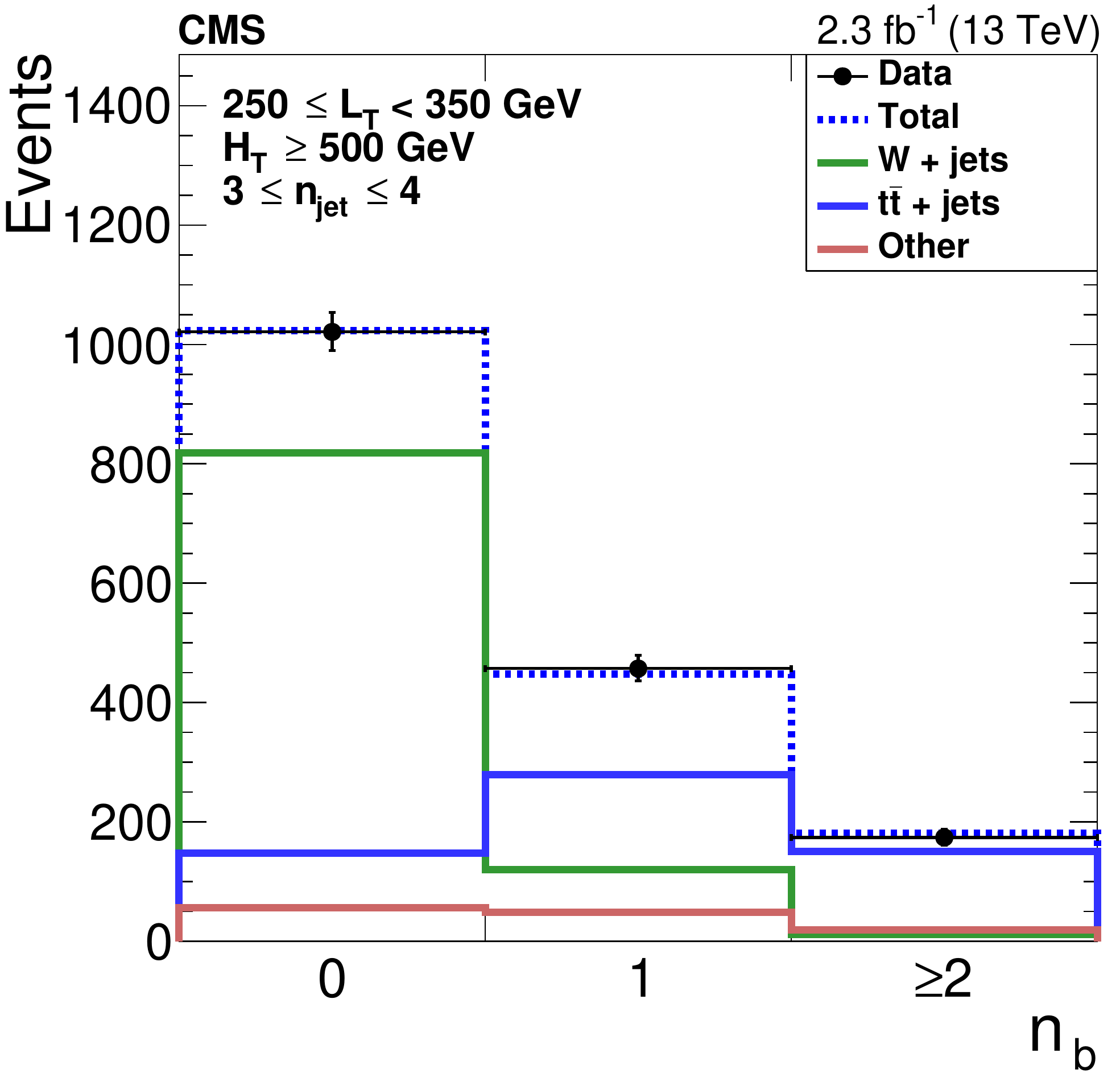}
    \includegraphics[width=0.48\textwidth]{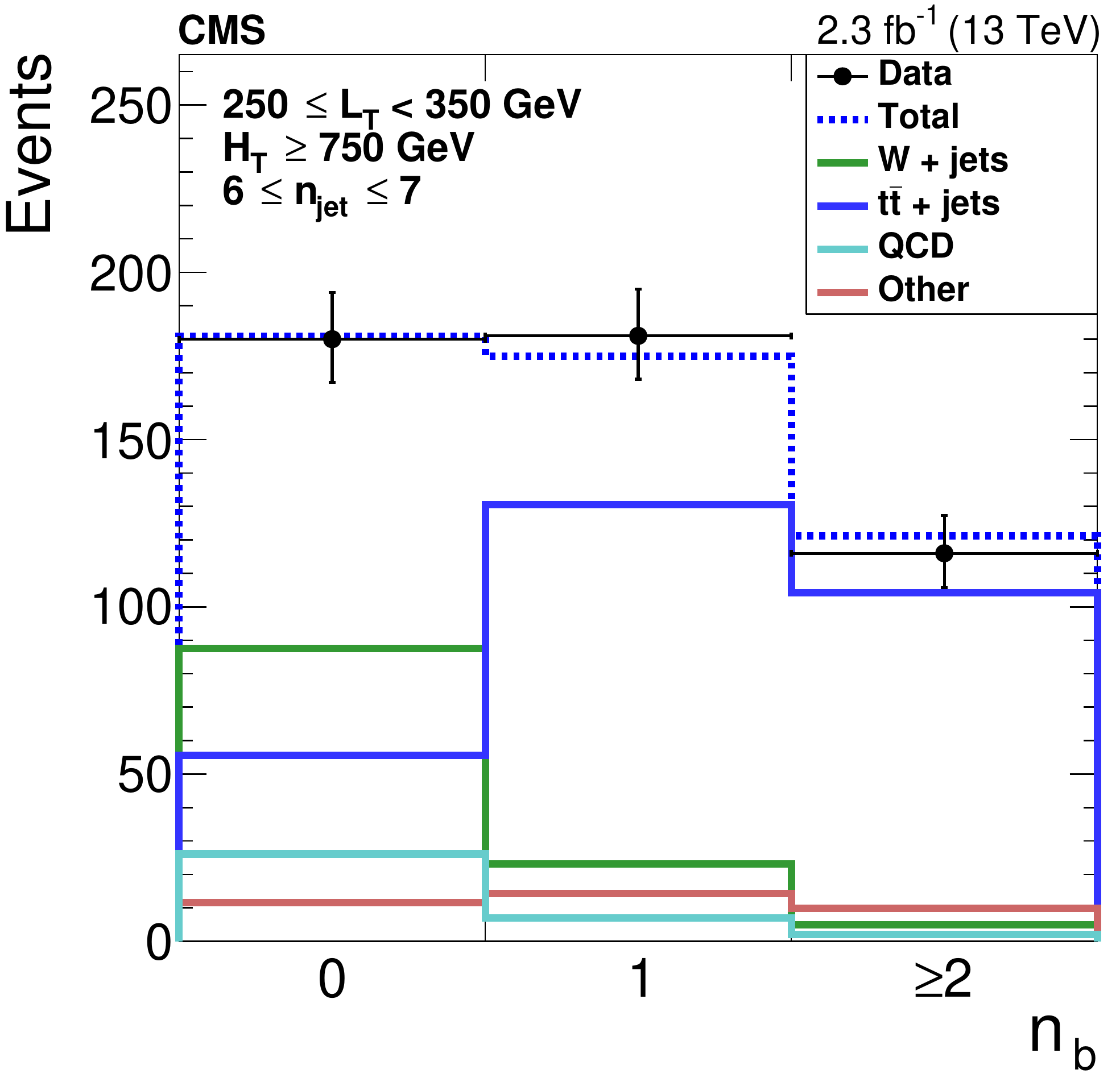}
    \caption{Fits to the \nbtag multiplicity for control regions in (\cmsLeft) $3\leq\njet\leq4$ ($250\leq$\LT$<350\GeV$, \HT$\geq500\GeV$, $\DF<1$) and (\cmsRight) $6\leq\njet\leq7$ ($250\leq$\LT$<350\GeV$, \HT$\geq750\GeV$, $\DF<1$) in data (muon channel).
The solid lines represent the templates scaled according to the fit result (blue for \ttjets, green for \Wjets, turquoise for QCD, and red for the remaining backgrounds), the dashed line shows the sum after fit, and the points with error bars represent data.}
    \label{fig:btagMultiplicityFit_data}
\end{figure}

The expected number of events for each of the background components can be described by:
\begin{equation}
N_i^\mathrm{SR} = N^\mathrm{CR}_\text{data} \, f_i \, \Rcs^{i},\; \text{ with }\; i=[\PW,\,\ttbar],
\end{equation}
where  $N^\mathrm{CR}_\text{data}$ is the total number of events in the CR of the MB region and $f_i$ is the relative yield of component $i$.
The relative contributions of the two components are determined by a fit of templates obtained from simulation to the \nbjet
multiplicity distribution in the CR of the MB region.
The contribution of the QCD multijet background in the CR is fixed to the yield estimated from data as described in Section~\ref{sec:qcdest}.
The contribution of other rare background components is obtained from simulation as well, as is done in the SR.
Uncertainties in these two components are propagated as systematic uncertainties to the final prediction.
Examples of these fits are shown in Fig.~\ref{fig:btagMultiplicityFit_data}.

The two \Rcs values, for \Wjets and \ttjets, are measured in two different low \njet SB regions.
For the \ttjets estimate a sideband with the requirements $4 \leq \njet \leq 5$ and $\nbtag=1$ is used.
The value of $\Rcs^{\ttbar}$ is then given by:
\begin{equation}
\Rcs^{\ttbar}(0\PQb,\njetSR) = \kappab\,\kappatt\,\Rcs^\text{data}(1\PQb,\,\njet \in [4,5]).
\end{equation}

The correction factors \kappab and \kappatt are determined from simulation.
The factor \kappab corrects for a potential difference of $\Rcs^{\ttbar}$ between samples with zero or one \PQb jet and for the small contributions of backgrounds other than \ttjets or QCD multijet events.
The factor \kappatt corrects for a residual dependence of $\Rcs^{\ttbar}$ on \njet, in analogy to the $\kappa_\mathrm{EW}$ factor defined in Section~\ref{sec:rcs_multib}.
Both values, \kappab and \kappatt, are close to unity, and statistical uncertainties from the simulation are propagated to the predicted yields.

Similarly, the \Wjets contribution is estimated using \Rcs values from a sideband with $3 \leq \njet \leq 4$ and $\nbjet = 0$.
With respect to the SB used for the estimate of $\Rcs^{\ttbar}$, a lower jet multiplicity is chosen in order to limit the contamination from \ttjets events.
Only the muon channel is used since it has a negligible contamination from QCD multijet events, contrary to the electron channel.
A systematic uncertainty is derived from simulation to cover potential differences between the $\mu$ and the combined \Pe\ and $\mu$ samples.
The value of $\Rcs^{\PW}$ is given by:
\begin{equation}
\Rcs^{\PW}(0\PQb,~\njetSR)=\kappaW\,\Rcscorr(0\PQb,\,\njet \in [3,4]).
\end{equation}

Again, the factor $\kappaW$ corrects for a residual dependence of $\Rcs^{\PW}$ on the jet multiplicity.
The raw value of $\Rcs^\text{data}$ measured in the SB has to be corrected for the contamination of \ttjets events.
The \ttjets yields are subtracted in the numerator and denominator according to:
\begin{equation}
\Rcs^{\text{data(corr)}}(0\PQb,\njet \in [3,4])=
\frac{N^{SR}_\text{data}-\Rcs^{\ttbar,\mathrm{MC}} \, f_{\ttbar} \, N^\mathrm{CR}_\text{data}}{(1-f_{\ttbar}) \, N^\mathrm{CR}_\text{data}}.
\end{equation}
The event yields $N^{CR}_\text{data}$ and $N^{SR}_\text{data}$ are measured in the SB CRs and SRs.
The fraction of \ttjets events $f_{\ttbar}$ is again obtained by a fit to the \nbjet multiplicity in the SB CR.
The \Rcs value for \ttjets in this SB is obtained from simulation.

Systematic uncertainties are assigned to $\kappatt$ and $\kappaW$ according to the difference between the \Rcs values in the sideband and the result of a linear fit over the full range of \njet.
The uncertainties vary from 3 to 43\% for \kappatt\ and from 1 to 49\% for \kappaW.
The two sources are treated as being independent.

\subsection{Estimate of the multijet background} \label{sec:qcdest}

Multijet events enter this analysis mostly when reconstructed electrons originate from misidentified
jets or from photon conversion in the inner detector. This background is estimated from the yield of `antiselected'
electron candidates in each region, that pass looser identification and isolation requirements, and fail the tighter criteria
for selected electrons. These events are scaled by the ratio of jets and photons that pass the tight
electron identification requirements to the number of antiselected electron candidates in a multijet-enriched control
sample with no \PQb-tagged jets and three or four other jets.
The assumption is that this sample is devoid of genuine prompt electrons.
The estimation method was introduced
previously~\cite{Chatrchyan:2011ig,Chatrchyan:2012ola}, and relies on the \Lp variable:
\begin{equation}
	\Lp = \frac{\pt^{\ell}}{\pt^{\PW}} \cos(\DF).
\end{equation}
For the dominant SM backgrounds, \ttjets and \Wjets, the distribution of \Lp is a well-understood consequence
of the $\PW$ boson polarization and falls from 0 to 1. In contrast, the distribution of \Lp for multijet events peaks
near $\Lp = 1$.

The ratio of selected to antiselected electron candidates is obtained from a fit to the \Lp distribution in bins of \LT.
The shape of the QCD multijet contribution used in the fit is taken from the antiselected sample, while the shape of all
other contributions is taken from simulation, as the behavior due to W polarization is well understood.
The ratios are found to be in the range 0.1--0.2.

In principle, the background estimation with the \Rcs method requires knowledge of the multijet contribution in the
SR and CR separately. Since the multijet background estimation is performed inclusively with respect to \DF, an \Rcs factor for
multijet events is determined as well.
In practice, since the resulting \Rcs values are all found to be below 2\%, the multijet contamination is
negligible for the SR. Therefore, the previously described \Rcs method takes into account only the QCD multijet contribution
in the CR, as written in Eq.~(\ref{eq:Rcs}).
For the muon channel, the contribution from QCD multijet background is typically of the order of 1\% of the total background.
To estimate this contribution, a procedure similar to the one outlined above is applied, and assigned a 100\% uncertainty.

\section{Systematic uncertainties} \label{sec:systematics}

Systematic uncertainties either influence $\kappa$, and thereby the predictions for the background, or modify
the expected signal yield.

The main systematic uncertainty on the background arises from the extrapolation of \Rcs from the low \njet region,
where it is measured, to the MB regions of higher jet multiplicities, where it is applied. Therefore, a systematic uncertainty
on \Rcs is determined in a dedicated control region with dileptonic events.
The ratio of the semileptonic to dileptonic
\ttjets final states for different numbers of reconstructed jets is of major importance, since the total \Rcs is based on
the fraction of the two channels and their corresponding \Rcs values, which differ significantly in \ttjets events.
To ensure that the data are described well by simulation, a high-purity dilepton \ttjets control sample
is selected from the data by requiring two leptons of opposite charge. For same-flavor
leptons it is also required that the invariant mass of the lepton pair be more than 10\GeV away from the \Z boson mass peak.
To study the behavior of the dileptonic events in the single-lepton selection, one of the two leptons
is removed from the event.
Since these ``lost leptons'' are principally from $\tau \to\text{hadrons} + \nu$ decays, we replace the removed
lepton by a jet with 2/3 of the
original lepton's \pt to accommodate for the missing energy due to the neutrino from the $\tau$ decay, and recalculate the \LT, \DF, and \HT values of the now ``single-lepton'' event.
In order to maximize the number of events, no \DF requirement is applied, and all events are used twice,
with each reconstructed lepton being considered as the lost lepton.
We refer to the samples produced using this procedure as the dilepton CRs.

A key test is performed by comparing the
jet multiplicity distribution in the sample resulting from single-lepton baseline selection (excluding the SRs) with
the corresponding simulated event sample, and by comparing the dilepton CRs with the corresponding simulated event
sample. Both comparisons show the same trend, a slight overprediction by simulation of the rate of high jet multiplicity events.
The ratio of event yields in data-to-simulation is computed for each comparison and the two ratios are then divided to
see whether the behavior in data relative to simulation is the same in both pairs of samples.
This double ratio is consistent
with unity within statistical uncertainty. The systematic uncertainty in the description of the \njet distribution in simulation
is determined from this double ratio, and is mainly due to the statistical uncertainty of the data samples, which is
within 8--40\%, and therefore larger than the observed slope of the double ratio \vs \njet.

The remaining uncertainties are smaller than the one from the dileptonic \ttjets fraction. In particular,
the applied jet energy scale (JES) factors are varied up and down according to their uncertainty~\cite{Chatrchyan:2011ds}
as a function of jet \pt and $\eta$, and these changes are propagated to \ETmiss.
The scale factors applied to the efficiencies for the identification of b-quark jets and for the misidentification of c-quark,
light-quark or gluon jets are also varied up and down according to their uncertainties~\cite{CMS-PAS-BTV-15-001}.
Uncertainties for the efficiency of lepton reconstruction and identification are handled in the same way.
For pileup, a 5\% uncertainty in the inelastic cross section~\cite{Aaboud:2016mmw} is used to obtain its impact on the uncertainty in the pileup.
In a few bins with low number of simulated events, the reweighting leads to a large uncertainty.
All these uncertainties apply to both the background prediction and the signal yield.
The luminosity is measured with the pixel cluster counting method, and the absolute luminosity scale calibration is derived from
an analysis of Van der Meer scans performed in August 2015, resulting in an uncertainty of 2.7\%~\cite{CMS-PAS-LUM-15-001}.

The \Wjets and \ttjets cross sections are changed by 30\%~\cite{Khachatryan:2016ipq} to cover possible biases in the estimation
of the background composition in terms of \Wjets \vs \ttjets events, which would lead to a slight change in the $\kappa$
value. These changes have only a small impact on the
zero-\PQb analysis, where the relative fraction of the two processes is determined from a fit.
Also, the following changes in the simulation are performed, with differences between the values of
$\kappa$ in the reweighted and original samples defining the uncertainties.
Motivated by measurements at $\sqrt{s} = 8\TeV$, simulated \ttjets events are reweighted by a factor $\sqrt{\smash[b]{F(\pt^{\PQt})\, F(\pt^{\PAQt})}}$, with $F(\pt^{\PQt}) = \min(0.5, \exp{(0.156 - 0.00137 \, \pt^{\PQt})})$, to improve the modelling of the top quark \pt spectrum~\cite{toppt_reweighting_8TeV}.
The reweighting preserves the normalization of the sample, and the difference relative to the results obtained with the unweighted sample is assigned as a systematic uncertainty.
The polarization of $\PW$ bosons is varied by reweighting events by the factor $w(\cos{\theta^*})= 1 + \alpha(1-\cos{\theta^*})^2$,
where $\theta^*$ is the angle between the charged lepton and $\PW$ boson in the $\PW$ boson rest frame.  In \Wjets events, we take
$\alpha$ to be 0.1, guided by the theoretical uncertainty and measurements found in Refs.~\cite{Bern:2011ie,Khachatryan:2015paa,Chatrchyan:2011ig,ATLAS:2012au}.
For \ttjets events, we take $\alpha= 0.05$.
For \Wjets events, where the initial state can have different polarizations for $\PWp$ \vs $\PWm$ bosons,
we take as uncertainty the larger change in $\kappa$ resulting from reweighting only the $\PWp$ bosons in the sample, and from reweighting all $\PW$ bosons.
The $\ttbar$V cross section is varied by 100\%.
The systematic uncertainty in the multijet estimation depends on \njet and \nbtag, and ranges from 25\% to 100\%.

For the zero-\PQb analysis, an additional systematic uncertainty is applied, based on
linear fits of \Rcs as a function of \njet as described in Section~\ref{sec:rcs_0b}, and
a 50\% cross section uncertainty is used for all backgrounds other than \Wjets, \ttjets, $\ttbar$V, and multijets.

For the signal, an uncertainty in initial-state radiation (ISR) is applied, based on the \pt of the gluino-gluino system,
which corresponds to a 15\% uncertainty at \pt between 400 and 600\GeV, and 30\% at larger \pt. This uncertainty is based
on measurements of ISR in $\Z$+jets and \ttjets events~\cite{CMS-PAS-SUS-15-007,Chatrchyan2013}. The factorization and
renormalization scale are each changed by a factor of 0.5 and 2.
Uncertainties in the signal cross section are also taken into account.

The impact of the systematic uncertainties in the total background prediction for the multi-\PQb and zero-\PQb analyses are summarized in Table~\ref{tab:sysTable}.
While the systematic uncertainty is determined for each signal point, the uncertainties typical
for most signals are summarized for illustration in Table~\ref{tab:sysSigTab}.

\begin{table*}[!htb]
\centering
\topcaption{Summary of systematic uncertainties in the total background prediction for the multi-\PQb and for the zero-\PQb analysis. }
\label{tab:sysTable}
\begin{scotch}{lcc}
Source     & Uncertainty for multi-\PQb [\%] & Uncertainty for zero-\PQb [\%]    \\ \hline
Dilepton control sample  & 5.8--20 & 7.5--40 \\
JES  & 0.2--11 & 0.6--8.2 \\
Tagging of \PQb-jets  & 0.1--17 & 1.4--4.5 \\
$\sigma(\Wjets)$ & 0.3--6.4 & $<$2.5 \\
W polarization  & 0.1--2 & 0.2--3.4 \\
$\sigma(\ttbar$V)  & 0.1--5 & 0.2--2.9 \\
Reweighting of top quark \pt  & 0.1--10 & 0.1--7.1 \\
Pileup  & 0.3--23 & 0.1--10 \\
Fit to \Rcs(\njet) (\Wjets and \ttjets) & \NA & 3.3--35 \\ \hline
Total & 8.0--28 & 10--54 \\ \hline
Statistical uncertainty in MC events & 3.0--30 & 8.2--48 \\
\end{scotch}
\end{table*}

\begin{table}[!h]
\centering
\topcaption{Summary of the systematic uncertainties and their average effect on the yields of the benchmark signals. The values are very similar for the multi-\PQb
and the zero-\PQb analysis, and are usually larger for compressed scenarios, where the mass difference between gluino and neutralino is small.}
\label{tab:sysSigTab}
\begin{scotch}{lc}

Source                              & Uncertainty [\%]     \\ \hline
Trigger                             & 1                      \\
Pileup                                 & 5                        \\
Lepton efficiency                   & 5                        \\
Luminosity                          & 2.7                      \\
ISR                                 & 3--20                     \\
Tagging of \PQb-jets (heavy flavors)      & 6--10                        \\
Tagging of \PQb-jets (light flavors)      & 2--3                        \\
JES                                 & 3--10                     \\
Factorization/renormalization scale & $<$ 3                         \\ \hline
Total                               & 12--26 \\
\end{scotch}
\end{table}
\section{Results and interpretation}\label{sec:results_interpretation}

\begin{figure*}[tbp!]
\centering
\includegraphics[width=0.75\textwidth]{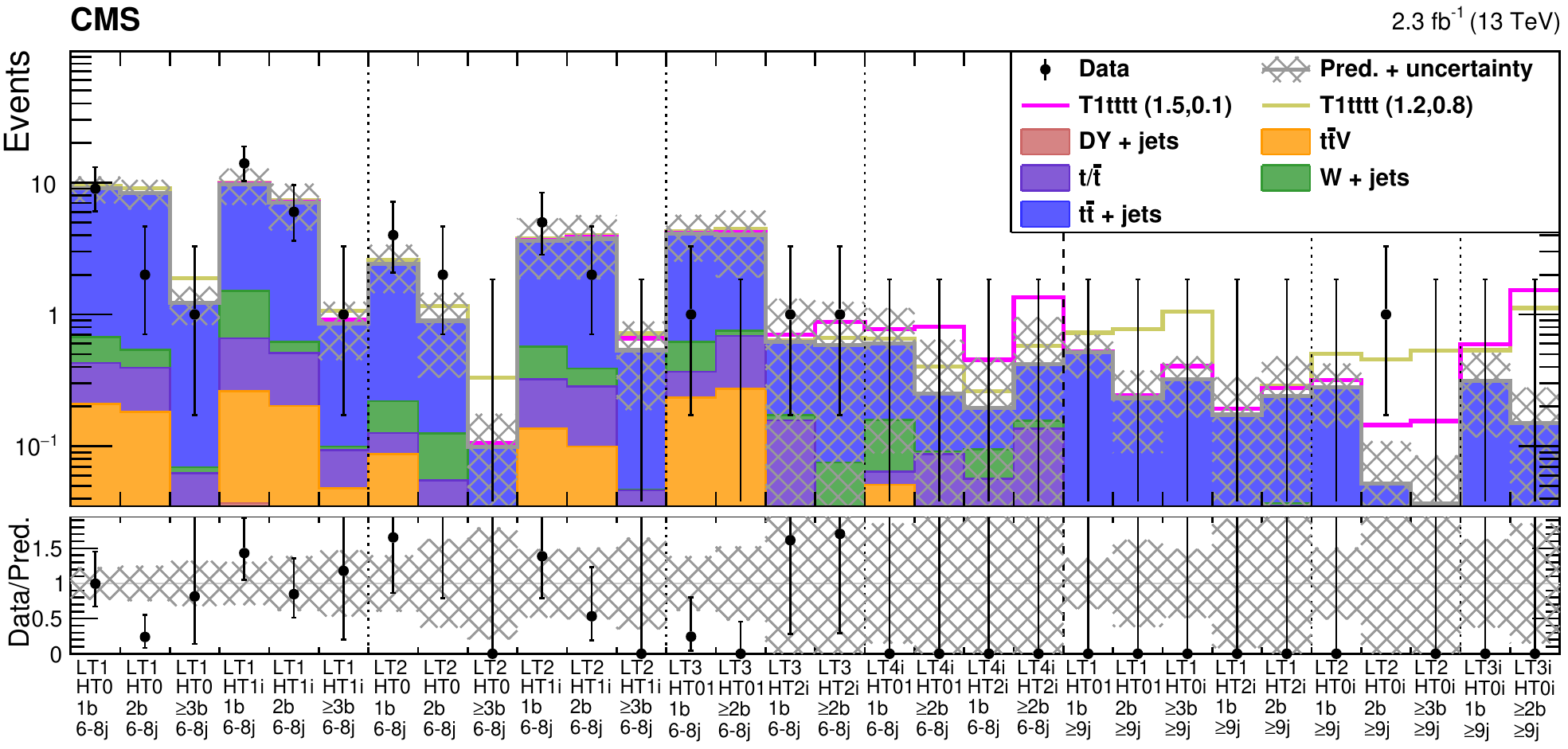}
\caption{
Multi-\PQb search: comparison of observed and predicted background event
yields in the 30 search regions. Upper panel:
the data are shown by black points with error bars, while the total SM
background predictions are shown by a grey line
with a hatched region representing its uncertainty. For illustration,
the relative fraction of the different SM background contributions, as
determined from simulation, is shown by the stacked, colored histograms,
whose total normalization is set by the
total background yields obtained from the control samples in the data.
The expected event yields for two T1tttt SUSY benchmark
models are shown by open histograms, each of which is shown stacked on
the total background prediction. The vertical dashed and dotted lines
separate different \njet and \LT bins, respectively, as indicated by the
$x$-axis labels. Lower panel: the ratio of the yield observed in data to
the predicted background yield is shown for each bin. The error bars on
the data points indicate the combined statistical and systematic
uncertainty in the ratio. The grey hatched region indicates the uncertainty on the
ratio that arises from the uncertainty on the background prediction.
}
\label{fig:multibunblind}
\end{figure*}

\begin{figure}[tbp!]
\centering
\includegraphics[width=\cmsFigWidth]{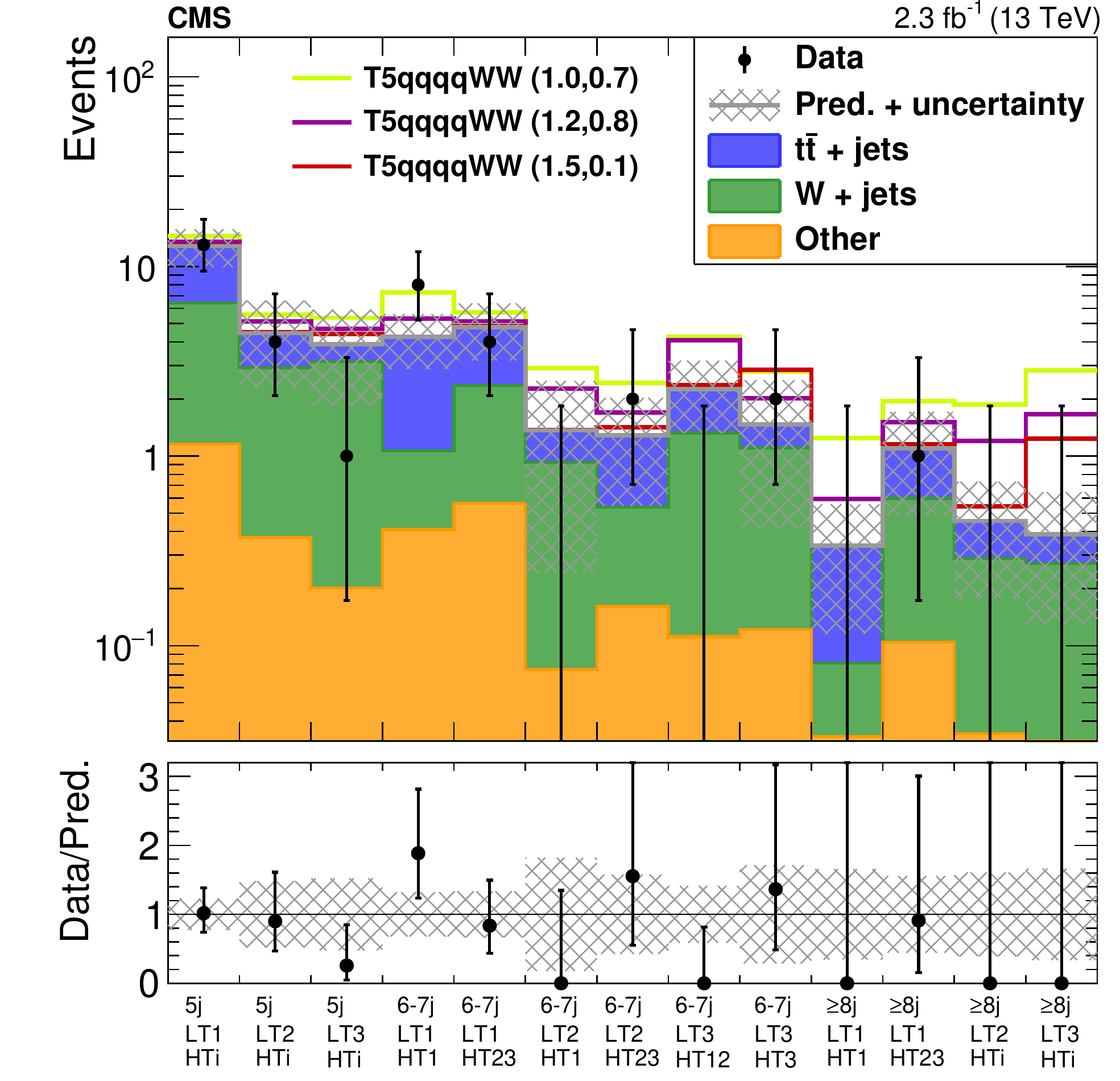}
\caption{Zero-\PQb search: observed and predicted event counts in the 13 search regions.
Upper panel: the black points with error bars show the number of observed events.
The filled, stacked histograms represent the predictions for \ttjets, \Wjets events, and the remaining backgrounds.
The uncertainty on the background prediction is shown as a grey, hatched region.
The expected yields from three T5qqqqWW model points, added to the SM background, are shown as solid lines.
Lower panel: the ratio of the yield observed in data to
the predicted background yield is shown for each bin. The error bars on
the data points indicate the combined statistical and systematic
uncertainty in the ratio. The grey hatched region indicates the uncertainty on the
ratio that arises from the uncertainty on the background prediction.
}
\label{fig:zerobunblind}
\end{figure}

The backgrounds for all SRs are determined, as described previously, in different SB regions with
lower jet or \PQb-jet multiplicities. The result of the background prediction and the observed data are shown in
Table~\ref{tab:PAS_table_multib} and Fig.~\ref{fig:multibunblind} for the multi-\PQb events. In this figure,
the outline of the filled histogram represents the total number of background events from the prediction.
For illustration, the relative amount of \ttjets, \Wjets, and of other backgrounds is shown as well, based on the
fractions estimated in simulation.
Table~\ref{tab:0b_resultTable} and Fig.~\ref{fig:zerobunblind} show the results for the zero-\PQb events. Here,
the filled histogram represents the predictions from data for \ttjets and \Wjets events, and for the remaining backgrounds,
where the latter include the multijet prediction determined from data and rare backgrounds taken from simulation.
The data agree with SM expectations and no excess is observed.

\begin{table*}[ht]
\topcaption{Summary of the results in the multi-\PQb search.}
\centering
\label{tab:PAS_table_multib}
\cmsTableResize{\begin{scotch}{c | c | c | c | l | c | c | c | c  }
\multirow{2}{*}{\njet} & \LT        &   \HT & \multirow{2}{*}{\nbjet} & \multirow{2}{*}{Bin name} &
\multicolumn{2}{c|}{Expected signal T1tttt $m_{\PSg}$/$m_{\PSGcz}$ $[$TeV$]$} & Predicted & \multirow{2}{*}{Observed}  \\
      & $[\GeVns{}]$  &   $[\GeVns{}]$ &    &          & \multicolumn{1}{c}{(1.5,0.1)}      &            \multicolumn{1}{c|}{(1.2,0.8)}         &            background                &   \\ \hline
\hline
  \multirow{20}{*}{[6, 8]} &  \multirow{6}{*}{[250, 350]} & \multirow{3}{*}{[500, 750]} & $=$ 1&LT1, HT0, NB1 & $<0.01$  & 0.41 \quad $\pm$  \quad 0.02  & 9.0  \quad $\pm$ \quad 2.1 & 9 \\
  & &  & $=$ 2&LT1, HT0, NB2 & $<0.01$ & 0.67  \quad $\pm$  \quad 0.03  & 8.4  \quad $\pm$ \quad 2.1 & 2 \\
  & &  & $\geq$3&LT1, HT0, NB3i & $<0.01$  & 0.67  \quad $\pm$  \quad 0.03  & 1.23  \quad $\pm$ \quad 0.39 & 1 \\
  \cline{3-9}  &  & \multirow{3}{*}{$\geq$750} & $=$ 1&LT1, HT1i, NB1 & 0.03  \quad $\pm$  \quad 0.00  & 0.15  \quad $\pm$  \quad 0.01  & 9.8  \quad $\pm$ \quad 3.0 & 14 \\
  & &  & $=$ 2&LT1, HT1i, NB2 & 0.07  \quad $\pm$  \quad 0.00  & 0.27  \quad $\pm$  \quad 0.02  & 7.1  \quad $\pm$ \quad 2.7 & 6 \\
  & &  & $\geq$3&LT1, HT1i, NB3i & 0.07  \quad $\pm$  \quad 0.00  & 0.22  \quad $\pm$  \quad 0.02  & 0.85  \quad $\pm$ \quad 0.40 & 1 \\
   \cline{2-9} &   \multirow{6}{*}{[350, 450]} & \multirow{3}{*}{[500, 750]} & $=$ 1&LT2, HT0, NB1 & $<0.01$  & 0.19  \quad $\pm$  \quad 0.02  & 2.42  \quad $\pm$ \quad 0.96 & 4 \\
  & &  & $=$ 2&LT2, HT0, NB2 & 0.01  \quad $\pm$  \quad 0.00  & 0.28  \quad $\pm$  \quad 0.02  & 0.89  \quad $\pm$ \quad 0.56 & 2 \\
  & &  & $\geq$3&LT2, HT0, NB3i & 0.01  \quad $\pm$  \quad 0.00  & 0.24  \quad $\pm$  \quad 0.02  & 0.10  \quad $\pm$ \quad 0.08 & 0 \\
  \cline{3-9}  &  & \multirow{3}{*}{$\geq$750} & $=$ 1&LT2, HT1i, NB1 & 0.08  \quad $\pm$  \quad 0.00  & 0.16  \quad $\pm$  \quad 0.01  & 3.6  \quad $\pm$ \quad 1.8 & 5 \\
  & &  & $=$ 2&LT2, HT1i, NB2 & 0.12  \quad $\pm$  \quad 0.01  & 0.24  \quad $\pm$  \quad 0.02  & 3.8  \quad $\pm$ \quad 1.9 & 2 \\
  & &  & $\geq$3&LT2, HT1i, NB3i & 0.13  \quad $\pm$  \quad 0.01  & 0.19  \quad $\pm$  \quad 0.01  & 0.54  \quad $\pm$ \quad 0.35 & 0 \\
   \cline{2-9} &   \multirow{4}{*}{[450, 600]} & \multirow{2}{*}{[500, 1250]} & $=$ 1&LT3, HT01, NB1 & 0.07  \quad $\pm$  \quad 0.00  & 0.18  \quad $\pm$  \quad 0.02  & 4.1  \quad $\pm$ \quad 1.6 & 1 \\
  & &  & $\geq$2&LT3, HT01, NB2i & 0.19  \quad $\pm$  \quad 0.01  & 0.42  \quad $\pm$  \quad 0.02  & 4.0  \quad $\pm$ \quad 2.1 & 0 \\
  \cline{3-9}  &  & \multirow{2}{*}{$\geq$1250} & $=$ 1&LT3, HT2i, NB1 & 0.08  \quad $\pm$  \quad 0.00  & 0.02  \quad $\pm$  \quad 0.00  & 0.62  \quad $\pm$ \quad 0.69 & 1 \\
  & &  & $\geq$2&LT3, HT2i, NB2i & 0.29  \quad $\pm$  \quad 0.01  & 0.08  \quad $\pm$  \quad 0.01  & 0.59  \quad $\pm$ \quad 0.66 & 1 \\
   \cline{2-9} &   \multirow{4}{*}{$\geq$600} & \multirow{2}{*}{[500, 1250]} & $=$ 1&LT4i, HT01, NB1 & 0.18  \quad $\pm$  \quad 0.01  & 0.05  \quad $\pm$  \quad 0.01  & 0.60  \quad $\pm$ \quad 0.51 & 0 \\
  & &  & $\geq$2&LT4i, HT01, NB2i & 0.57  \quad $\pm$  \quad 0.01  & 0.16  \quad $\pm$  \quad 0.01  & 0.25  \quad $\pm$ \quad 0.39 & 0 \\
  \cline{3-9}  &  & \multirow{2}{*}{$\geq$1250} & $=$ 1&LT4i, HT2i, NB1 & 0.26  \quad $\pm$  \quad 0.01  & 0.07  \quad $\pm$  \quad 0.01  & 0.20  \quad $\pm$ \quad 0.27 & 0 \\
  & &  & $\geq$2&LT4i, HT2i, NB2i & 0.95  \quad $\pm$  \quad 0.02  & 0.16  \quad $\pm$  \quad 0.01  & 0.42  \quad $\pm$ \quad 0.53 & 0 \\
 \hline
\hline
   \multirow{10}{*}{$\geq$9}  &   \multirow{5}{*}{[250, 350]} & \multirow{2}{*}{[500, 1250]} & $=$ 1&LT1, HT01, NB1 & 0.01  \quad $\pm$  \quad 0.00  & 0.22  \quad $\pm$  \quad 0.02  & 0.52  \quad $\pm$ \quad 0.19 & 0 \\
  & &  & $=$ 2&LT1, HT01, NB2 & 0.01  \quad $\pm$  \quad 0.00  & 0.55  \quad $\pm$  \quad 0.03  & 0.23  \quad $\pm$ \quad 0.14 & 0 \\
  \cline{3-9}  &  & $\geq$500 & $\geq$3&LT1, HT0i, NB3i & 0.08  \quad $\pm$  \quad 0.00  & 0.74  \quad $\pm$  \quad 0.03  & 0.32  \quad $\pm$ \quad 0.16 & 0 \\
  \cline{3-9}  &  & \multirow{2}{*}{$\geq$1250} & $=$ 1&LT1, HT2i, NB1 & 0.02  \quad $\pm$  \quad 0.00  & 0.02  \quad $\pm$  \quad 0.01  & 0.17  \quad $\pm$ \quad 0.16 & 0 \\
  & &  & $=$ 2&LT1, HT2i, NB2 & 0.04  \quad $\pm$  \quad 0.00  & 0.05  \quad $\pm$  \quad 0.01  & 0.24  \quad  $\pm$ \quad 0.31 & 0 \\
   \cline{2-9} &   \multirow{3}{*}{[350, 450]} & \multirow{3}{*}{$\geq$500} & $=$ 1&LT2, HT0i, NB1 & 0.04  \quad $\pm$  \quad 0.00  & 0.23  \quad $\pm$  \quad 0.02  & 0.28  \quad $\pm$ \quad 0.14 & 0 \\
  & &  & $=$ 2&LT2, HT0i, NB2 & 0.10  \quad $\pm$  \quad 0.01  & 0.41  \quad $\pm$  \quad 0.02  & 0.05  \quad $\pm$ \quad 0.06 & 1 \\
  & &  & $\geq$3&LT2, HT0i, NB3i & 0.12  \quad $\pm$  \quad 0.01  & 0.51  \quad $\pm$  \quad 0.02  & 0.04  \quad $\pm$ \quad 0.05 & 0 \\
   \cline{2-9} &   \multirow{2}{*}{$\geq$450} & \multirow{2}{*}{$\geq$500} & $=$ 1&LT3i, HT0i, NB1 & 0.29  \quad $\pm$  \quad 0.01  & 0.23  \quad $\pm$  \quad 0.02  & 0.31  \quad $\pm$ \quad 0.20 & 0 \\
  & &  & $\geq$2&LT3i, HT0i, NB2i & 1.42  \quad $\pm$  \quad 0.02  & 0.99  \quad $\pm$  \quad 0.03  & 0.15  \quad $\pm$ \quad 0.13 & 0 \\
\end{scotch}}
\end{table*}

\begin{table*}[htbp]
\centering
\topcaption{
  Summary of the results of the zero-\PQb search.
}
\label{tab:0b_resultTable}
\resizebox{\textwidth}{!}{\begin{scotch}{c|c|c|l|rrr|rrr|rrr|rrr|rrr}
\multirow{2}{*}{\njet}     & \LT & \HT     & \multirow{2}{*}{Bin name} & \multicolumn{9}{c|}{Expected signal T5qqqqWW $m_{\PSg}$/$m_{\PSGcz}$ $[$TeV$]$} & \multicolumn{3}{c|}{Predicted} & \multicolumn{3}{c}{\multirow{2}{*}{Observed}} \\
 & $[\GeVns{}]$ &$[\GeVns{}]$ &  & \multicolumn{3}{c}{(1.0,0.7)} & \multicolumn{3}{c}{(1.2,0.8)} & \multicolumn{3}{c|}{(1.5,0.1)} & \multicolumn{3}{c|}{background} & \multicolumn{3}{c}{} \\\hline
\hline
\multirow{3}{*}{5}
&\multirow{1}{*}{$[250,350]$}
&${\geq}$500
 & $\mathrm{LT1}, \mathrm{HTi}$
 & 1.67&$\pm$&0.27 & 0.68&$\pm$&0.07 & 0.03&$\pm$&0.01 & 12.8&$\pm$&2.9
 & \multicolumn{3}{c}{13} \\
\cline{2-19}
&\multirow{1}{*}{$[350,450]$}
&${\geq}$500
 & $\mathrm{LT2}, \mathrm{HTi}$
 & 1.13&$\pm$&0.22 & 0.68&$\pm$&0.07 & 0.04&$\pm$&0.01 & 4.5&$\pm$&2.2
 & \multicolumn{3}{c}{4} \\
\cline{2-19}
&\multirow{1}{*}{${\geq}$450}
&${\geq}$500
 & $\mathrm{LT3}, \mathrm{HTi}$
 & 1.48&$\pm$&0.26 & 0.79&$\pm$&0.08 & 0.51&$\pm$&0.02 & 3.9&$\pm$&2.0
 & \multicolumn{3}{c}{1} \\
\cline{2-19}
\hline
\hline
\multirow{6}{*}{[6,7]}
&\multirow{2}{*}{$[250,350]$}
&$[500,750]$
 & $\mathrm{LT1}, \mathrm{HT1}$
 & 3.03&$\pm$&0.36 & 1.06&$\pm$&0.09 & \multicolumn{3}{c|}{$<0.01$} & 4.2 &$\pm$&1.4
 & \multicolumn{3}{c}{8} \\
&
&${\geq}$750
 & $\mathrm{LT1}, \mathrm{HT23}$
 & 0.92&$\pm$&0.20 & 0.36&$\pm$&0.05 & 0.08&$\pm$&0.01 & 4.8&$\pm$&1.6
 & \multicolumn{3}{c}{4} \\
\cline{2-19}
&\multirow{2}{*}{$[350,450]$}
&$[500,750]$
 & $\mathrm{LT2}, \mathrm{HT1}$
 & 1.54&$\pm$&0.26 & 0.90&$\pm$&0.08 & \multicolumn{3}{c|}{$<0.01$} & 1.4&$\pm$&1.1
 & \multicolumn{3}{c}{0} \\
&
&${\geq}$750
 & $\mathrm{LT2}, \mathrm{HT23}$
 & 1.15&$\pm$&0.21 & 0.41&$\pm$&0.05 & 0.13&$\pm$&0.01 & 1.29&$\pm$&0.74
 & \multicolumn{3}{c}{2} \\
\cline{2-19}
&\multirow{2}{*}{${\geq}$450}
&$[500,1000]$
 & $\mathrm{LT3}, \mathrm{HT12}$
 & 1.99&$\pm$&0.29 & 1.83&$\pm$&0.12 & 0.11&$\pm$&0.01 & 2.25&$\pm$&0.93
 & \multicolumn{3}{c}{0} \\
&
&${\geq}$1000
 & $\mathrm{LT3}, \mathrm{HT3}$
 & 1.33&$\pm$&0.23 & 0.55&$\pm$&0.06 & 1.38&$\pm$&0.04 & 1.5&$\pm$&1.0
 & \multicolumn{3}{c}{2} \\
\cline{2-19}
\hline
\hline
\multirow{4}{*}{${\geq}$8}
&\multirow{2}{*}{$[250,350]$}
&$[500,750]$
 & $\mathrm{LT1}, \mathrm{HT1}$
 & 0.90&$\pm$&0.20 & 0.26&$\pm$&0.04 & \multicolumn{3}{c|}{$<0.01$} & 0.34&$\pm$&0.22
 & \multicolumn{3}{c}{0} \\
&
&${\geq}$750
 & $\mathrm{LT1}, \mathrm{HT23}$
 & 0.85&$\pm$&0.19 & 0.41&$\pm$&0.05 & 0.06&$\pm$&0.01 & 1.10&$\pm$&0.61
 & \multicolumn{3}{c}{1} \\
\cline{2-19}
&\multirow{1}{*}{$[350,450]$}
&${\geq}$500
 & $\mathrm{LT2}, \mathrm{HTi}$
 & 1.41&$\pm$&0.23 & 0.75&$\pm$&0.07 & 0.09&$\pm$&0.01 & 0.45&$\pm$&0.28
 & \multicolumn{3}{c}{0} \\
\cline{2-19}
&\multirow{1}{*}{${\geq}$450}
&${\geq}$500
 & $\mathrm{LT3}, \mathrm{HTi}$
 & 2.44&$\pm$&0.31 & 1.27&$\pm$&0.09 & 0.84&$\pm$&0.03 & 0.39&$\pm$&0.26
 & \multicolumn{3}{c}{0} \\
\end{scotch}}
\end{table*}

To set limits, separate likelihood functions, one for the multi-\PQb analysis and one for the zero-\PQb analysis,
are constructed from the Poisson probability functions for all four
data regions (the CRs and SRs in the SB as well in the MB) to determine the background in the MB SR.
In addition, the $\kappa$ values from simulation are included to correct any residual differences between the SB
and MB regions, with uncertainties incorporated through log-normal constraints.
The estimated contribution from multijet events in the two CRs is also included.
A possible signal contamination is taken into account by including signal terms in the fit for both the sideband and the control regions.
For the zero-\PQb analysis, the relative contributions of \Wjets and \ttjets events as determined in the fits to the \nbtag distribution in the CRs are treated as external measurements.
The correlation between the \Wjets and \ttjets yields introduced by these fits is taken into account.
A profile likelihood ratio in the asymptotic
approximation~\cite{Cowan:2010js} is used as the test statistic. Limits
are then calculated at the 95\% confidence level (\CL) using the
asymptotic CL$_\mathrm{s}$ criterion~\cite{Junk1999,ClsCite}.

\begin{figure}
\centering
\includegraphics[width=0.48\textwidth]{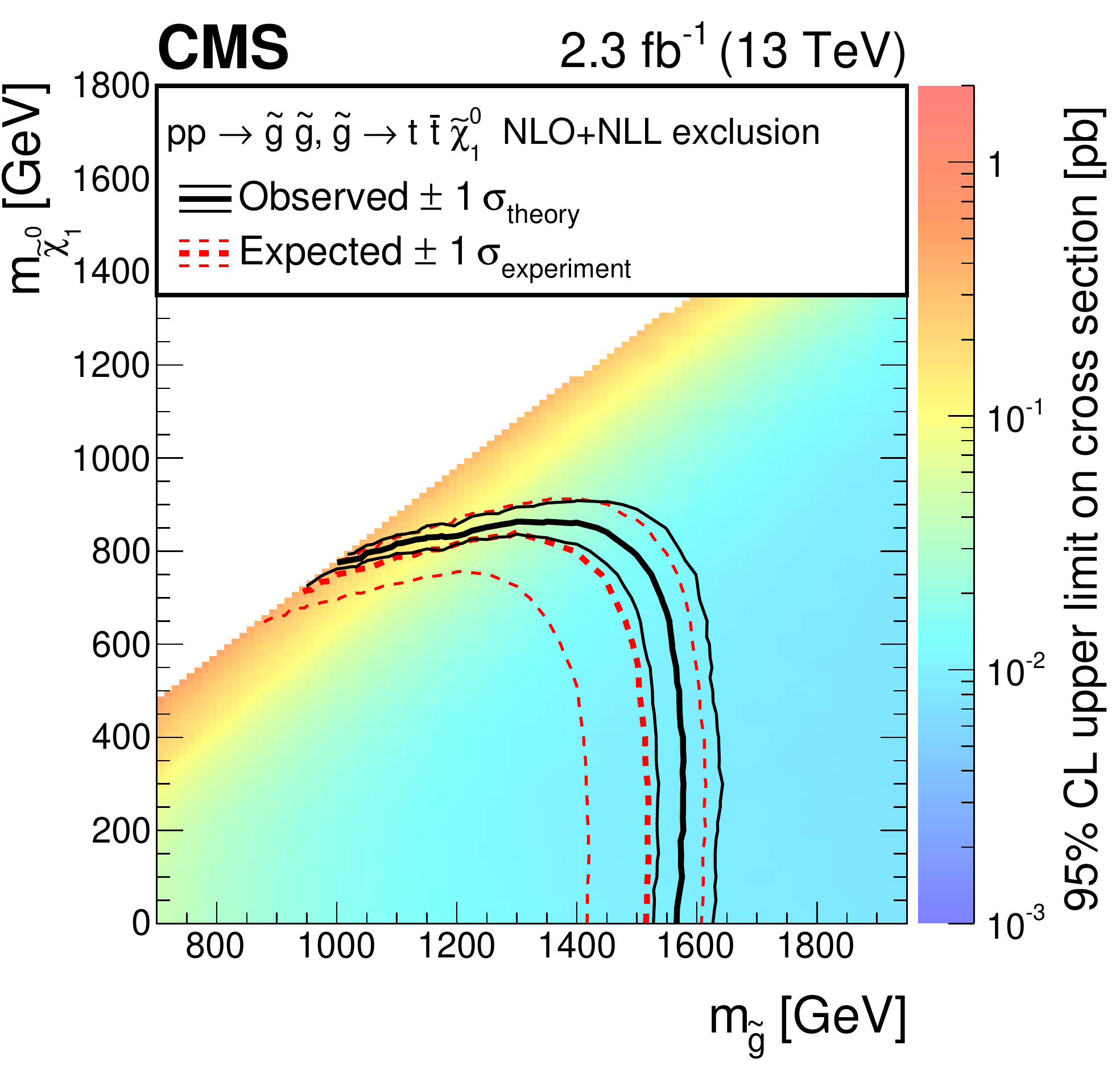} \hfil
\includegraphics[width=0.48\textwidth]{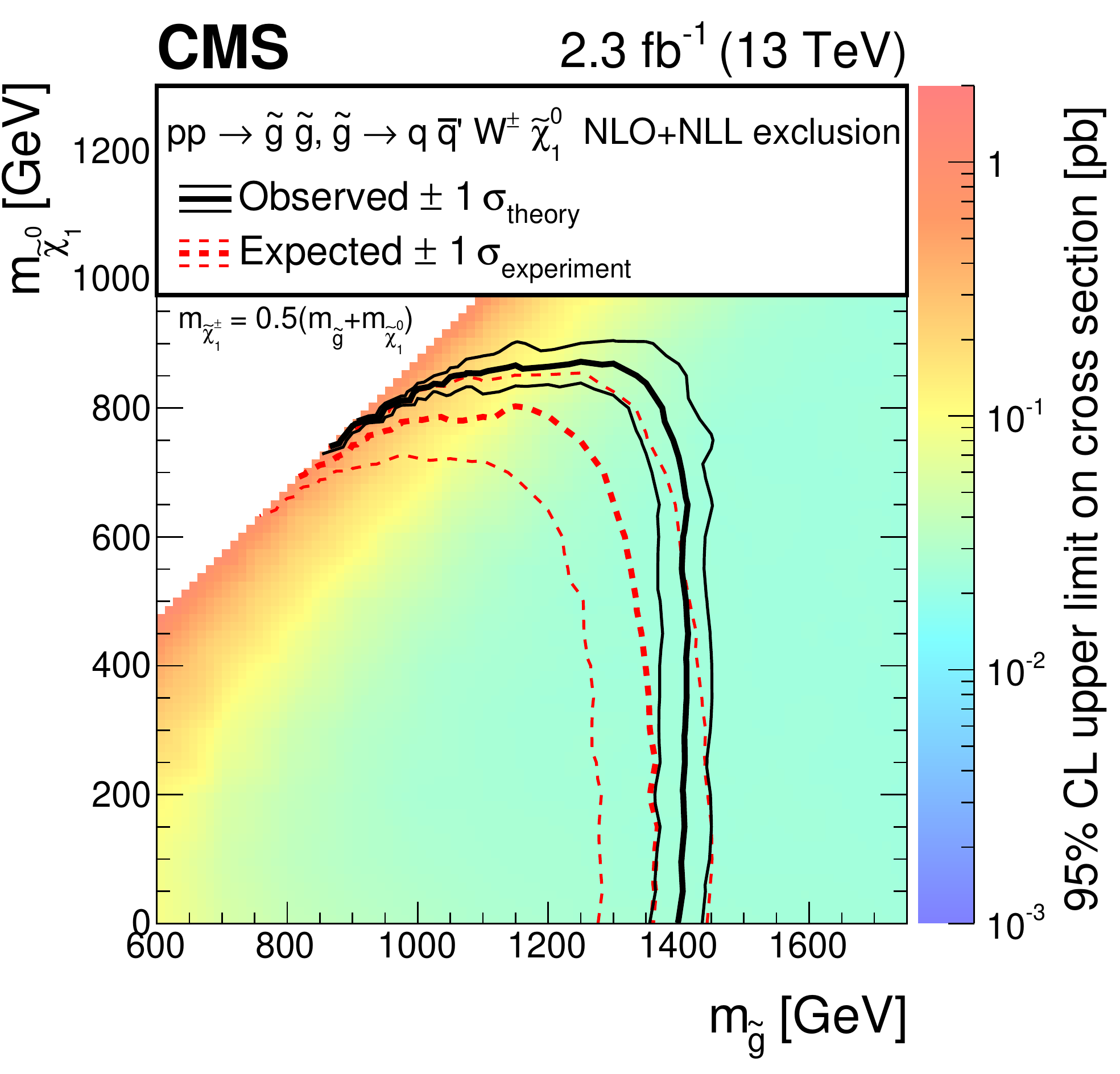}
\caption{Cross section limits at a 95\% \CL for the (\cmsLeft) T1tttt and (\cmsRight) T5qqqqWW models, as a function of the gluino and LSP masses.
In T5qqqqWW, the pair-produced
gluinos each decay to a quark-antiquark pair of the first or second generation (\qqbar), and a chargino (\PSGcpmDo) with its mass taken to be $m_{\PSGcpmDo}=0.5(m_{\PSg}+m_{\PSGczDo})$.
The solid black (dashed red) lines correspond to the observed (expected) mass limits, with the thicker lines representing the central values and the thinner lines representing the ${\pm}1\sigma$ uncertainty bands related to the theoretical (experimental) uncertainties.}\label{fig:limits}
\end{figure}

The cross section limits obtained for the T1tttt model using the multi-\PQb analysis, and for the T5qqqqWW model using the zero-\PQb analysis, are shown in
Fig.~\ref{fig:limits} as a function of $m(\PSg)$ and $m(\PSGczDo)$, assuming branching fractions of 100\% as shown in Fig.~\ref{fig:feynman_1}.
Using the \PSg\PSg\ pair production cross section calculated at next-to-leading order within next-to-leading-logarithmic accuracy, exclusion limits are set as a function
of the $(m_{\PSg},m_{\PSGczDo})$ mass hypothesis.

\section{Summary} \label{sec:summary}

A search for supersymmetry has been performed with 2.3\fbinv of proton-proton collision data recorded by the CMS experiment at $\sqrt{s}=13\TeV$ in 2015.
The data are analyzed in several exclusive categories, differing in the number of jets and \PQb-tagged jets, the scalar sum
of all jet transverse momenta, and the scalar sum of the missing transverse momentum and the transverse momentum
of the lepton. The main background is significantly reduced
by requiring a large azimuthal angle between the directions of the momenta of the lepton and of the reconstructed $\PW$ boson.
No significant excess is observed, and the results are interpreted in terms of two simplified models that
describe gluino pair production.

For the simplified model T1tttt, in which each gluino decays through an off-shell top squark to a \ttbar pair and the lightest neutralino, gluino masses up to 1.6\TeV are excluded for neutralino masses below 600\GeV. Neutralino masses below 850\GeV
can be excluded for a gluino mass up to 1.4\TeV.
Similar to Ref.~\cite{CMS-PAS-SUS-15-007}, these results extend the limits obtained from the 8\TeV searches~\cite{Chatrchyan:2013iqa,Aad:2015mia,Aad:2014lra} by about 250\GeV.

The second simplified model T5qqqqWW also contains gluino pair production, with the gluinos decaying to first or second
generation squarks and a chargino, which then decays to a $\PW$ boson and the lightest neutralino.
The chargino mass in this decay chain is taken to be $m_{\PSGcpmDo}=0.5(m_{\PSg}+m_{\PSGczDo})$.
In this model, gluino masses below 1.4\TeV are excluded for neutralino masses below 700\GeV. For a gluino mass of 1.3\TeV, neutralinos with masses up to 850\GeV can be excluded.
These results improve existing limits~\cite{ATLAS-13TeV_single_lepton} on the neutralino mass in this channel for gluino masses between 900\GeV and 1.4\TeV.

\begin{acknowledgments}
We congratulate our colleagues in the CERN accelerator departments for the excellent performance of the LHC and thank the technical and administrative staffs at CERN and at other CMS institutes for their contributions to the success of the CMS effort. In addition, we gratefully acknowledge the computing centers and personnel of the Worldwide LHC Computing Grid for delivering so effectively the computing infrastructure essential to our analyses. Finally, we acknowledge the enduring support for the construction and operation of the LHC and the CMS detector provided by the following funding agencies: BMWFW and FWF (Austria); FNRS and FWO (Belgium); CNPq, CAPES, FAPERJ, and FAPESP (Brazil); MES (Bulgaria); CERN; CAS, MoST, and NSFC (China); COLCIENCIAS (Colombia); MSES and CSF (Croatia); RPF (Cyprus); SENESCYT (Ecuador); MoER, ERC IUT and ERDF (Estonia); Academy of Finland, MEC, and HIP (Finland); CEA and CNRS/IN2P3 (France); BMBF, DFG, and HGF (Germany); GSRT (Greece); OTKA and NIH (Hungary); DAE and DST (India); IPM (Iran); SFI (Ireland); INFN (Italy); MSIP and NRF (Republic of Korea); LAS (Lithuania); MOE and UM (Malaysia); BUAP, CINVESTAV, CONACYT, LNS, SEP, and UASLP-FAI (Mexico); MBIE (New Zealand); PAEC (Pakistan); MSHE and NSC (Poland); FCT (Portugal); JINR (Dubna); MON, RosAtom, RAS and RFBR (Russia); MESTD (Serbia); SEIDI and CPAN (Spain); Swiss Funding Agencies (Switzerland); MST (Taipei); ThEPCenter, IPST, STAR and NSTDA (Thailand); TUBITAK and TAEK (Turkey); NASU and SFFR (Ukraine); STFC (United Kingdom); DOE and NSF (USA).

\hyphenation{Rachada-pisek} Individuals have received support from the Marie-Curie program and the European Research Council and EPLANET (European Union); the Leventis Foundation; the A. P. Sloan Foundation; the Alexander von Humboldt Foundation; the Belgian Federal Science Policy Office; the Fonds pour la Formation \`a la Recherche dans l'Industrie et dans l'Agriculture (FRIA-Belgium); the Agentschap voor Innovatie door Wetenschap en Technologie (IWT-Belgium); the Ministry of Education, Youth and Sports (MEYS) of the Czech Republic; the Council of Science and Industrial Research, India; the HOMING PLUS program of the Foundation for Polish Science, cofinanced from European Union, Regional Development Fund, the Mobility Plus program of the Ministry of Science and Higher Education, the National Science Center (Poland), contracts Harmonia 2014/14/M/ST2/00428, Opus 2013/11/B/ST2/04202, 2014/13/B/ST2/02543 and 2014/15/B/ST2/03998, Sonata-bis 2012/07/E/ST2/01406; the Thalis and Aristeia programs cofinanced by EU-ESF and the Greek NSRF; the National Priorities Research Program by Qatar National Research Fund; the Programa Clar\'in-COFUND del Principado de Asturias; the Rachadapisek Sompot Fund for Postdoctoral Fellowship, Chulalongkorn University and the Chulalongkorn Academic into Its 2nd Century Project Advancement Project (Thailand); and the Welch Foundation, contract C-1845. \end{acknowledgments}

\clearpage
\bibliography{auto_generated}

\cleardoublepage \appendix\section{The CMS Collaboration \label{app:collab}}\begin{sloppypar}\hyphenpenalty=5000\widowpenalty=500\clubpenalty=5000\input{SUS-15-006-authorlist.tex}\end{sloppypar}
\end{document}

%% file: SUS-15-006-authorlist.tex
\textbf{Yerevan Physics Institute,  Yerevan,  Armenia}\\*[0pt]
V.~Khachatryan, A.M.~Sirunyan, A.~Tumasyan
\vskip\cmsinstskip
\textbf{Institut f\"{u}r Hochenergiephysik der OeAW,  Wien,  Austria}\\*[0pt]
W.~Adam, E.~Asilar, T.~Bergauer, J.~Brandstetter, E.~Brondolin, M.~Dragicevic, J.~Er\"{o}, M.~Flechl, M.~Friedl, R.~Fr\"{u}hwirth\cmsAuthorMark{1}, V.M.~Ghete, C.~Hartl, N.~H\"{o}rmann, J.~Hrubec, M.~Jeitler\cmsAuthorMark{1}, A.~K\"{o}nig, I.~Kr\"{a}tschmer, D.~Liko, T.~Matsushita, I.~Mikulec, D.~Rabady, N.~Rad, B.~Rahbaran, H.~Rohringer, J.~Schieck\cmsAuthorMark{1}, J.~Strauss, W.~Treberer-Treberspurg, W.~Waltenberger, C.-E.~Wulz\cmsAuthorMark{1}
% ...
\vskip\cmsinstskip
% ...
1:~~Also at Vienna University of Technology, Vienna, Austria\\
% ...


%% file: SUS-15-006_temp.bbl
\providecommand{\href}[2]{#2}\begingroup\raggedright\begin{thebibliography}{10}%
\makeatletter
\providecommand{\hrefCMSnoop }[0]{\@secondoftwo}%
\makeatother
\providecommand{\doi}{\texttt{doi:}\begingroup \urlstyle{tt}\Url}

\bibitem{Ramond:1971gb}
\hrefCMSnoop {}{P.~Ramond, ``{Dual theory for free fermions}'',} \textit{ Phys.
  Rev. D} \textbf{ 3} (1971) 2415,
\href{http://dx.doi.org/10.1103/PhysRevD.3.2415}{\doi{10.1103/PhysRevD.3.2415}}.

\bibitem{Golfand:1971iw}
\hrefCMSnoop {}{Y.~A. Golfand and E.~P. Likhtman, ``{Extension of the algebra
  of {P}oincar\'{e} group generators and violation of {P} invariance}'',}
  \textit{ JETP Lett.} \textbf{ 13} (1971)
323.

\bibitem{Neveu:1971rx}
\hrefCMSnoop {}{A.~Neveu and J.~H. Schwarz, ``{Factorizable dual model of
  pions}'',} \textit{ Nucl. Phys. B} \textbf{ 31} (1971) 86,
\href{http://dx.doi.org/10.1016/0550-3213(71)90448-2}{\doi{10.1016/0550-3213(71)90448-2}}.

\bibitem{Volkov:1972jx}
\hrefCMSnoop {}{D.~V. Volkov and V.~P. Akulov, ``{Possible universal neutrino
  interaction}'',} \textit{ JETP Lett.} \textbf{ 16} (1972)
438.

\bibitem{Wess:1973kz}
\hrefCMSnoop {}{J.~Wess and B.~Zumino, ``{A {L}agrangian model invariant under
  supergauge transformations}'',} \textit{ Phys. Lett. B} \textbf{ 49} (1974)
  52,
\href{http://dx.doi.org/10.1016/0370-2693(74)90578-4}{\doi{10.1016/0370-2693(74)90578-4}}.

\bibitem{Wess:1974tw}
\hrefCMSnoop {}{J.~Wess and B.~Zumino, ``{Supergauge transformations in four
  dimensions}'',} \textit{ Nucl. Phys. B} \textbf{ 70} (1974) 39,
\href{http://dx.doi.org/10.1016/0550-3213(74)90355-1}{\doi{10.1016/0550-3213(74)90355-1}}.

\bibitem{Fayet:1974pd}
\hrefCMSnoop {}{P.~Fayet, ``{Supergauge invariant extension of the {H}iggs
  mechanism and a model for the electron and its neutrino}'',} \textit{ Nucl.
  Phys. B} \textbf{ 90} (1975) 104,
\href{http://dx.doi.org/10.1016/0550-3213(75)90636-7}{\doi{10.1016/0550-3213(75)90636-7}}.

\bibitem{Nilles:1983ge}
\hrefCMSnoop {}{H.~P. Nilles, ``{Supersymmetry, supergravity and particle
  physics}'',} \textit{ Phys. Rep.} \textbf{ 110} (1984) 1,
\href{http://dx.doi.org/10.1016/0370-1573(84)90008-5}{\doi{10.1016/0370-1573(84)90008-5}}.

\bibitem{FARRAR1978575}
\hrefCMSnoop {}{G.~R. Farrar and P.~Fayet, ``Phenomenology of the production,
  decay, and detection of new hadronic states associated with supersymmetry'',}
  \textit{ Phys. Lett. B} \textbf{ 76} (1978) 575,
  \href{http://dx.doi.org/10.1016/0370-2693(78)90858-4}{\doi{10.1016/0370-2693(78)90858-4}}.

\bibitem{Chatrchyan:2012ola}
\hrefCMSnoop {}{{CMS Collaboration}, ``{Search for supersymmetry in pp
  collisions at $\sqrt{s}=7\TeV$ in events with a single lepton, jets, and
  missing transverse momentum}'',} \textit{ Eur. Phys. J. C} \textbf{ 73}
  (2013) 2404,
  \href{http://dx.doi.org/10.1140/epjc/s10052-013-2404-z}{\doi{10.1140/epjc/s10052-013-2404-z}},
\href{http://www.arXiv.org/abs/1212.6428}{\texttt{arXiv:1212.6428}}.

\bibitem{Chatrchyan:2012sca}
\hrefCMSnoop {}{{CMS Collaboration}, ``{Search for supersymmetry in final
  states with a single lepton, b-quark jets, and missing transverse energy in
  proton-proton collisions at $\sqrt{s}=7\TeV$}'',} \textit{ Phys. Rev. D}
  \textbf{ 87} (2013) 052006,
  \href{http://dx.doi.org/10.1103/PhysRevD.87.052006}{\doi{10.1103/PhysRevD.87.052006}},
\href{http://www.arXiv.org/abs/1211.3143}{\texttt{arXiv:1211.3143}}.

\bibitem{Aad:2012ms}
\hrefCMSnoop {}{{ATLAS Collaboration}, ``{Further search for supersymmetry at
  $\sqrt{s}=7$\TeV in final states with jets, missing transverse momentum and
  isolated leptons with the ATLAS detector}'',} \textit{ Phys. Rev. D} \textbf{
  86} (2012) 092002,
  \href{http://dx.doi.org/10.1103/PhysRevD.86.092002}{\doi{10.1103/PhysRevD.86.092002}},
\href{http://www.arXiv.org/abs/1208.4688}{\texttt{arXiv:1208.4688}}.

\bibitem{Chatrchyan:2013iqa}
\hrefCMSnoop {}{{CMS Collaboration}, ``{Search for supersymmetry in pp
  collisions at $\sqrt{s}=8\TeV$ in events with a single lepton, large jet
  multiplicity, and multiple b jets}'',} \textit{ Phys. Lett. B} \textbf{ 733}
  (2014) 328,
  \href{http://dx.doi.org/10.1016/j.physletb.2014.04.023}{\doi{10.1016/j.physletb.2014.04.023}},
\href{http://www.arXiv.org/abs/1311.4937}{\texttt{arXiv:1311.4937}}.

\bibitem{Aad:2015mia}
\hrefCMSnoop {}{{ATLAS Collaboration}, ``{Search for squarks and gluinos in
  events with isolated leptons, jets and missing transverse momentum at
  $\sqrt{s}=8$ TeV with the ATLAS detector}'',} \textit{ JHEP} \textbf{ 04}
  (2015) 116,
  \href{http://dx.doi.org/10.1007/JHEP04(2015)116}{\doi{10.1007/JHEP04(2015)116}},
\href{http://www.arXiv.org/abs/1501.03555}{\texttt{arXiv:1501.03555}}.

\bibitem{Aad:2014lra}
\hrefCMSnoop {}{{ATLAS Collaboration}, ``{Search for strong production of
  supersymmetric particles in final states with missing transverse momentum and
  at least three b-jets at $\sqrt{s} = 8\TeV$ proton-proton collisions with the
  ATLAS detector}'',} \textit{ JHEP} \textbf{ 10} (2014) 024,
  \href{http://dx.doi.org/10.1007/JHEP10(2014)024}{\doi{10.1007/JHEP10(2014)024}},
\href{http://www.arXiv.org/abs/1407.0600}{\texttt{arXiv:1407.0600}}.

\bibitem{CMS-PAS-SUS-15-007}
\hrefCMSnoop {}{{CMS Collaboration}, ``{Search for supersymmetry in pp
  collisions at $\sqrt{s}=13$ TeV in the single-lepton final state using the
  sum of masses of large-radius jets}'',} \textit{ JHEP} \textbf{ 08} (2016)
  122,
  \href{http://dx.doi.org/10.1007/JHEP08(2016)122}{\doi{10.1007/JHEP08(2016)122}},
\href{http://www.arXiv.org/abs/1605.04608}{\texttt{arXiv:1605.04608}}.

\bibitem{ATLAS-13TeV_single_lepton}
\hrefCMSnoop {}{{ATLAS Collaboration}, ``{Search for gluinos in events with an
  isolated lepton, jets and missing transverse momentum at $\sqrt{s}$ = 13 TeV
  with the ATLAS detector}'',} (2016).
  \href{http://www.arXiv.org/abs/1605.04285}{\texttt{arXiv:1605.04285}}.
Submitted to \textit{EPJC}.

\bibitem{ATLAS-13TeV_multib}
\hrefCMSnoop {}{{ATLAS Collaboration}, ``{Search for pair production of gluinos
  decaying via stop and sbottom in events with $b$-jets and large missing
  transverse momentum in $pp$ collisions at $\sqrt{s} = 13$ TeV with the ATLAS
  detector}'',} (2016).
\href{http://www.arXiv.org/abs/1605.09318}{\texttt{arXiv:1605.09318}}.

\bibitem{bib-sms-1}
N.~Arkani-Hamed\hrefCMSnoop {}{ {et~al.}, ``{{MARMOSET}: The path from {LHC}
  data to the new standard model via on-shell effective theories}'',} (2007).
\href{http://www.arXiv.org/abs/hep-ph/0703088}{\texttt{arXiv:hep-ph/0703088}}.

\bibitem{bib-sms-2}
\hrefCMSnoop {}{J.~Alwall, P.~C. Schuster, and N.~Toro, ``Simplified models for
  a first characterization of new physics at the {LHC}'',} \textit{ Phys. Rev.
  D} \textbf{ 79} (2009) 075020,
  \href{http://dx.doi.org/10.1103/PhysRevD.79.075020}{\doi{10.1103/PhysRevD.79.075020}},
\href{http://www.arXiv.org/abs/0810.3921}{\texttt{arXiv:0810.3921}}.

\bibitem{bib-sms-3}
\hrefCMSnoop {}{J.~Alwall, M.-P. Le, M.~Lisanti, and J.~G. Wacker,
  ``{Model-independent jets plus missing energy searches}'',} \textit{ Phys.
  Rev. D} \textbf{ 79} (2009) 015005,
  \href{http://dx.doi.org/10.1103/PhysRevD.79.015005}{\doi{10.1103/PhysRevD.79.015005}},
\href{http://www.arXiv.org/abs/0809.3264}{\texttt{arXiv:0809.3264}}.

\bibitem{bib-sms-4}
D.~Alves\hrefCMSnoop {}{ {et~al.}, ``Simplified models for {LHC} new physics
  searches'',} \textit{ J. Phys. G} \textbf{ 39} (2012) 105005,
  \href{http://dx.doi.org/10.1088/0954-3899/39/10/105005}{\doi{10.1088/0954-3899/39/10/105005}},
\href{http://www.arXiv.org/abs/1105.2838}{\texttt{arXiv:1105.2838}}.

\bibitem{Chatrchyan:2008zzk}
\hrefCMSnoop {}{{CMS Collaboration}, ``The {CMS} experiment at the {CERN}
  {LHC}'',} \textit{ JINST} \textbf{ 3} (2008) S08004,
\href{http://dx.doi.org/10.1088/1748-0221/3/08/S08004}{\doi{10.1088/1748-0221/3/08/S08004}}.

\bibitem{TRK-11-001}
\hrefCMSnoop {}{{CMS Collaboration}, ``{Description and performance of track
  and primary-vertex reconstruction with the CMS tracker}'',} \textit{ JINST}
  \textbf{ 9} (2014) P10009,
  \href{http://dx.doi.org/10.1088/1748-0221/9/10/P10009}{\doi{10.1088/1748-0221/9/10/P10009}},
\href{http://www.arXiv.org/abs/1405.6569}{\texttt{arXiv:1405.6569}}.

\bibitem{Chatrchyan:2011ds}
\hrefCMSnoop {}{{CMS Collaboration}, ``{Determination of jet energy calibration
  and transverse momentum resolution in {CMS}}'',} \textit{ JINST} \textbf{ 6}
  (2011) P11002,
  \href{http://dx.doi.org/10.1088/1748-0221/6/11/P11002}{\doi{10.1088/1748-0221/6/11/P11002}},
\href{http://www.arXiv.org/abs/1107.4277}{\texttt{arXiv:1107.4277}}.

\bibitem{Khachatryan:2015hwa}
\hrefCMSnoop {}{{CMS Collaboration}, ``{Performance of electron reconstruction
  and selection with the CMS detector in proton-proton collisions at $\sqrt{s}
  = 8$\TeV}'',} \textit{ JINST} \textbf{ 10} (2015) P06005,
  \href{http://dx.doi.org/10.1088/1748-0221/10/06/P06005}{\doi{10.1088/1748-0221/10/06/P06005}},
\href{http://www.arXiv.org/abs/1502.02701}{\texttt{arXiv:1502.02701}}.

\bibitem{Chatrchyan:2012xi}
\hrefCMSnoop {}{{CMS Collaboration}, ``{Performance of CMS muon reconstruction
  in pp collision events at $\sqrt{s} = 7$\TeV}'',} \textit{ JINST} \textbf{ 7}
  (2012) P10002,
  \href{http://dx.doi.org/10.1088/1748-0221/7/10/P10002}{\doi{10.1088/1748-0221/7/10/P10002}},
\href{http://www.arXiv.org/abs/1206.4071}{\texttt{arXiv:1206.4071}}.

\bibitem{CMS-PAS-PFT-09-001}
\href {http://cdsweb.cern.ch/record/1194487}{{CMS Collaboration},
  ``{Particle-flow event reconstruction in {CMS} and performance for jets,
  taus, and {\MET}}'',} CMS Physics Analysis Summary CMS-PAS-PFT-09-001, CERN,
  2009.

\bibitem{CMS-PAS-PFT-10-001}
\href {http://cdsweb.cern.ch/record/1247373}{{CMS Collaboration},
  ``{Commissioning of the particle-flow event with the first {LHC} collisions
  recorded in the {CMS} detector}'',} CMS Physics Analysis Summary
  CMS-PAS-PFT-10-001, CERN, 2010.

\bibitem{Cacciari:2008gp}
\hrefCMSnoop {}{M.~Cacciari, G.~P. Salam, and G.~Soyez, ``{The Anti-k(t) jet
  clustering algorithm}'',} \textit{ JHEP} \textbf{ 04} (2008) 063,
  \href{http://dx.doi.org/10.1088/1126-6708/2008/04/063}{\doi{10.1088/1126-6708/2008/04/063}},
\href{http://www.arXiv.org/abs/0802.1189}{\texttt{arXiv:0802.1189}}.

\bibitem{Cacciari:2011ma}
\hrefCMSnoop {}{M.~Cacciari, G.~P. Salam, and G.~Soyez, ``{FastJet User
  Manual}'',} \textit{ Eur. Phys. J.} \textbf{ C72} (2012) 1896,
  \href{http://dx.doi.org/10.1140/epjc/s10052-012-1896-2}{\doi{10.1140/epjc/s10052-012-1896-2}},
\href{http://www.arXiv.org/abs/1111.6097}{\texttt{arXiv:1111.6097}}.

\bibitem{Cacciari:2007fd}
\hrefCMSnoop {}{M.~Cacciari and G.~P. Salam, ``{Pileup subtraction using jet
  areas}'',} \textit{ Phys. Lett. B} \textbf{ 659} (2008) 119,
  \href{http://dx.doi.org/10.1016/j.physletb.2007.09.077}{\doi{10.1016/j.physletb.2007.09.077}},
\href{http://www.arXiv.org/abs/0707.1378}{\texttt{arXiv:0707.1378}}.

\bibitem{Chatrchyan:2012jua}
\hrefCMSnoop {}{{CMS Collaboration}, ``{Identification of \cPqb-quark jets with
  the CMS experiment}'',} \textit{ JINST} \textbf{ 8} (2013) P04013,
  \href{http://dx.doi.org/10.1088/1748-0221/8/04/P04013}{\doi{10.1088/1748-0221/8/04/P04013}},
\href{http://www.arXiv.org/abs/1211.4462}{\texttt{arXiv:1211.4462}}.

\bibitem{CMS-PAS-BTV-15-001}
\href {http://cds.cern.ch/record/2138504}{{{CMS}} Collaboration,
  ``Identification of b quark jets at the CMS Experiment in the LHC Run 2'',}
  CMS Physics Analysis Summary CMS-PAS-BTV-15-001, CERN, 2016.

\bibitem{CMS-PAS-BTV-13-001}
\href {http://cds.cern.ch/record/1581306}{{{CMS}} Collaboration, ``Performance
  of $\cPqb$ tagging at $\sqrt{s}=8\TeV$ in multijet, \ttbar and boosted
  topology events'',} CMS Physics Analysis Summary CMS-PAS-BTV-13-001, CERN,
  2013.

\bibitem{Alwall:2011uj}
J.~Alwall\hrefCMSnoop {}{ {et~al.}, ``{M}ad{G}raph5: going beyond'',} \textit{
  JHEP} \textbf{ 06} (2011) 128,
  \href{http://dx.doi.org/10.1007/JHEP06(2011)128}{\doi{10.1007/JHEP06(2011)128}},
\href{http://www.arXiv.org/abs/1106.0522}{\texttt{arXiv:1106.0522}}.

\bibitem{Ball:2014uwa}
\hrefCMSnoop {}{{NNPDF} Collaboration, ``{Parton distributions for the LHC Run
  II}'',} \textit{ JHEP} \textbf{ 04} (2015) 040,
  \href{http://dx.doi.org/10.1007/JHEP04(2015)040}{\doi{10.1007/JHEP04(2015)040}},
\href{http://www.arXiv.org/abs/1410.8849}{\texttt{arXiv:1410.8849}}.

\bibitem{Nason:2004rx}
\hrefCMSnoop {}{P.~Nason, ``{A new method for combining NLO QCD with shower
  Monte Carlo algorithms}'',} \textit{ JHEP} \textbf{ 11} (2004) 040,
  \href{http://dx.doi.org/10.1088/1126-6708/2004/11/040}{\doi{10.1088/1126-6708/2004/11/040}},
\href{http://www.arXiv.org/abs/hep-ph/0409146}{\texttt{arXiv:hep-ph/0409146}}.

\bibitem{Frixione:2007vw}
\hrefCMSnoop {}{S.~Frixione, P.~Nason, and C.~Oleari, ``{Matching NLO QCD
  computations with parton shower simulations: the POWHEG method}'',} \textit{
  JHEP} \textbf{ 11} (2007) 070,
  \href{http://dx.doi.org/10.1088/1126-6708/2007/11/070}{\doi{10.1088/1126-6708/2007/11/070}},
\href{http://www.arXiv.org/abs/0709.2092}{\texttt{arXiv:0709.2092}}.

\bibitem{Alioli:2010xd}
\hrefCMSnoop {}{S.~Alioli, P.~Nason, C.~Oleari, and E.~Re, ``{A general
  framework for implementing NLO calculations in shower Monte Carlo programs:
  the POWHEG BOX}'',} \textit{ JHEP} \textbf{ 06} (2010) 043,
  \href{http://dx.doi.org/10.1007/JHEP06(2010)043}{\doi{10.1007/JHEP06(2010)043}},
\href{http://www.arXiv.org/abs/1002.2581}{\texttt{arXiv:1002.2581}}.

\bibitem{Alioli:2009je}
\hrefCMSnoop {}{S.~Alioli, P.~Nason, C.~Oleari, and E.~Re, ``{NLO single-top
  production matched with shower in POWHEG: $s$- and $t$-channel
  contributions}'',} \textit{ JHEP} \textbf{ 09} (2009) 111,
  \href{http://dx.doi.org/10.1088/1126-6708/2009/09/111}{\doi{10.1088/1126-6708/2009/09/111}},
  \href{http://www.arXiv.org/abs/0907.4076}{\texttt{arXiv:0907.4076}}.
[Erratum: \DOI{10.1007/JHEP02(2010)011}].

\bibitem{Re:2010bp}
\hrefCMSnoop {}{E.~Re, ``{Single-top Wt-channel production matched with parton
  showers using the POWHEG method}'',} \textit{ Eur. Phys. J. C} \textbf{ 71}
  (2011) 1547,
  \href{http://dx.doi.org/10.1140/epjc/s10052-011-1547-z}{\doi{10.1140/epjc/s10052-011-1547-z}},
\href{http://www.arXiv.org/abs/1009.2450}{\texttt{arXiv:1009.2450}}.

\bibitem{Alwall:2014hca}
J.~Alwall\hrefCMSnoop {}{ {et~al.}, ``{The automated computation of tree-level
  and next-to-leading order differential cross sections, and their matching to
  parton shower simulations}'',} \textit{ JHEP} \textbf{ 07} (2014) 079,
  \href{http://dx.doi.org/10.1007/JHEP07(2014)079}{\doi{10.1007/JHEP07(2014)079}},
\href{http://www.arXiv.org/abs/1405.0301}{\texttt{arXiv:1405.0301}}.

\bibitem{Sjostrand:2014zea}
T.~Sj{\"o}strand\hrefCMSnoop {}{ {et~al.}, ``{An Introduction to PYTHIA
  8.2}'',} \textit{ Comput. Phys. Commun.} \textbf{ 191} (2015) 159,
  \href{http://dx.doi.org/10.1016/j.cpc.2015.01.024}{\doi{10.1016/j.cpc.2015.01.024}},
\href{http://www.arXiv.org/abs/1410.3012}{\texttt{arXiv:1410.3012}}.

\bibitem{bib-nlo-nll-01}
\hrefCMSnoop {}{W.~Beenakker, R.~H{\"o}pker, M.~Spira, and P.~M. Zerwas,
  ``Squark and gluino production at hadron colliders'',} \textit{ Nucl. Phys.
  B} \textbf{ 492} (1997) 51,
  \href{http://dx.doi.org/10.1016/S0550-3213(97)00084-9}{\doi{10.1016/S0550-3213(97)00084-9}},
\href{http://www.arXiv.org/abs/hep-ph/9610490}{\texttt{arXiv:hep-ph/9610490}}.

\bibitem{bib-nlo-nll-02}
\hrefCMSnoop {}{A.~Kulesza and L.~Motyka, ``{Threshold resummation for
  squark-antisquark and gluino-pair production at the {LHC}}'',} \textit{ Phys.
  Rev. Lett.} \textbf{ 102} (2009) 111802,
  \href{http://dx.doi.org/10.1103/PhysRevLett.102.111802}{\doi{10.1103/PhysRevLett.102.111802}},
\href{http://www.arXiv.org/abs/0807.2405}{\texttt{arXiv:0807.2405}}.

\bibitem{bib-nlo-nll-03}
\hrefCMSnoop {}{A.~Kulesza and L.~Motyka, ``{Soft gluon resummation for the
  production of gluino-gluino and squark-antisquark pairs at the {LHC}}'',}
  \textit{ Phys. Rev. D} \textbf{ 80} (2009) 095004,
  \href{http://dx.doi.org/10.1103/PhysRevD.80.095004}{\doi{10.1103/PhysRevD.80.095004}},
\href{http://www.arXiv.org/abs/0905.4749}{\texttt{arXiv:0905.4749}}.

\bibitem{bib-nlo-nll-04}
W.~Beenakker\hrefCMSnoop {}{ {et~al.}, ``{Soft-gluon resummation for squark and
  gluino hadroproduction}'',} \textit{ JHEP} \textbf{ 12} (2009) 041,
  \href{http://dx.doi.org/10.1088/1126-6708/2009/12/041}{\doi{10.1088/1126-6708/2009/12/041}},
\href{http://www.arXiv.org/abs/0909.4418}{\texttt{arXiv:0909.4418}}.

\bibitem{bib-nlo-nll-05}
W.~Beenakker\hrefCMSnoop {}{ {et~al.}, ``{Squark and gluino
  hadroproduction}'',} \textit{ Int. J. Mod. Phys. A} \textbf{ 26} (2011) 2637,
  \href{http://dx.doi.org/10.1142/S0217751X11053560}{\doi{10.1142/S0217751X11053560}},
\href{http://www.arXiv.org/abs/1105.1110}{\texttt{arXiv:1105.1110}}.

\bibitem{Agostinelli:2002hh}
\hrefCMSnoop {}{S.~Agostinelli {et~al.}, ``{GEANT4} --- a simulation
  toolkit'',} \textit{ Nucl. Instr. and Meth. A} \textbf{ 506} (2003) 250,
\href{http://dx.doi.org/10.1016/S0168-9002(03)01368-8}{\doi{10.1016/S0168-9002(03)01368-8}}.

\bibitem{bib-cms-fastsim-02}
\hrefCMSnoop {}{{CMS Collaboration}, ``The Fast Simulation of the CMS Detector
  at LHC'',} in \textit{ Int'l Conf. on Computing in High Energy and Nuclear
  Physics (CHEP 2010)}.
\newblock 2011.
\newblock {Journal of Physics: Conference Series, 331 (2011), 032049}.
  \href{http://dx.doi.org/10.1088/1742-6596/331/3/032049}{\doi{10.1088/1742-6596/331/3/032049}}.

\bibitem{Chatrchyan:2011ig}
\hrefCMSnoop {}{{CMS Collaboration}, ``{Measurement of the Polarization of W
  Bosons with Large Transverse Momenta in W+Jets Events at the LHC}'',}
  \textit{ Phys. Rev. Lett.} \textbf{ 107} (2011) 021802,
  \href{http://dx.doi.org/10.1103/PhysRevLett.107.021802}{\doi{10.1103/PhysRevLett.107.021802}},
\href{http://www.arXiv.org/abs/1104.3829}{\texttt{arXiv:1104.3829}}.

\bibitem{Aaboud:2016mmw}
\hrefCMSnoop {}{{ATLAS Collaboration}, ``{Measurement of the Inelastic
  Proton-Proton Cross Section at $\sqrt{s} = 13$ TeV with the ATLAS Detector at
  the LHC}'',} (2016).
  \href{http://www.arXiv.org/abs/1606.02625}{\texttt{arXiv:1606.02625}}.
Submitted to \textit{PRL}.

\bibitem{CMS-PAS-LUM-15-001}
\href {http://cdsweb.cern.ch/record/2138682}{{CMS Collaboration}, ``CMS
  Luminosity Measurement for the 2015 Data Taking Period'',} CMS Physics
  Analysis Summary CMS-PAS-LUM-15-001, CERN, 2016.

\bibitem{Khachatryan:2016ipq}
\hrefCMSnoop {}{{CMS Collaboration}, ``{Measurement of the production cross
  section of the W boson in association with two b jets in pp collisions at
  $\sqrt{s}= 8$\TeV}'',} (2016).
  \href{http://www.arXiv.org/abs/1608.07561}{\texttt{arXiv:1608.07561}}.
Submitted to {EPJC}.

\bibitem{toppt_reweighting_8TeV}
\hrefCMSnoop {}{{CMS Collaboration}, ``{Measurement of the differential cross
  section for top quark pair production in pp collisions at
  $\sqrt{s}=8\TeV$}'',} \textit{ Eur. Phys. J. C} \textbf{ 75} (2015) 542,
  \href{http://dx.doi.org/10.1140/epjc/s10052-015-3709-x}{\doi{10.1140/epjc/s10052-015-3709-x}},
\href{http://www.arXiv.org/abs/1505.04480}{\texttt{arXiv:1505.04480}}.

\bibitem{Bern:2011ie}
Z.~Bern\hrefCMSnoop {}{ {et~al.}, ``{Left-handed W bosons at the LHC}'',}
  \textit{ Phys. Rev. D} \textbf{ 84} (2011) 034008,
  \href{http://dx.doi.org/10.1103/PhysRevD.84.034008}{\doi{10.1103/PhysRevD.84.034008}},
\href{http://www.arXiv.org/abs/1103.5445}{\texttt{arXiv:1103.5445}}.

\bibitem{Khachatryan:2015paa}
\hrefCMSnoop {}{{CMS Collaboration}, ``{Angular coefficients of Z bosons
  produced in pp collisions at $\sqrt{s}= 8\TeV$ and decaying to $\mu^+ \mu^-$
  as a function of transverse momentum and rapidity}'',} \textit{ Phys. Lett.
  B} \textbf{ 750} (2015) 154,
  \href{http://dx.doi.org/10.1016/j.physletb.2015.08.061}{\doi{10.1016/j.physletb.2015.08.061}},
\href{http://www.arXiv.org/abs/1504.03512}{\texttt{arXiv:1504.03512}}.

\bibitem{ATLAS:2012au}
\hrefCMSnoop {}{{ATLAS Collaboration}, ``{Measurement of the polarisation of W
  bosons produced with large transverse momentum in pp collisions at
  $\sqrt{s}=7$ TeV with the ATLAS experiment}'',} \textit{ Eur. Phys. J. C}
  \textbf{ 72} (2012) 2001,
  \href{http://dx.doi.org/10.1140/epjc/s10052-012-2001-6}{\doi{10.1140/epjc/s10052-012-2001-6}},
\href{http://www.arXiv.org/abs/1203.2165}{\texttt{arXiv:1203.2165}}.

\bibitem{Chatrchyan2013}
\hrefCMSnoop {}{{CMS Collaboration}, ``Search for top-squark pair production in
  the single-lepton final state in pp collisions at {$\sqrt{s}=8$\TeV}'',}
  \textit{ Eur. Phys. J. C} \textbf{ 73} (2013) 2677,
  \href{http://dx.doi.org/10.1140/epjc/s10052-013-2677-2}{\doi{10.1140/epjc/s10052-013-2677-2}}.

\bibitem{Cowan:2010js}
\hrefCMSnoop {}{G.~Cowan, K.~Cranmer, E.~Gross, and O.~Vitells, ``Asymptotic
  formulae for likelihood-based tests of new physics'',} \textit{ Eur. Phys. J.
  C} \textbf{ 71} (2011) 1554,
  \href{http://dx.doi.org/10.1140/epjc/s10052-011-1554-0}{\doi{10.1140/epjc/s10052-011-1554-0}},
  \href{http://www.arXiv.org/abs/1007.1727}{\texttt{arXiv:1007.1727}}.
[Erratum: \DOI{10.1140/epjc/s10052-013-2501-z}].

\bibitem{Junk1999}
\hrefCMSnoop {}{T.~Junk, ``{Confidence level computation for combining searches
  with small statistics}'',} \textit{ Nucl. Instr. and Meth. A} \textbf{ 434}
  (1999) 435,
  \href{http://dx.doi.org/10.1016/S0168-9002(99)00498-2}{\doi{10.1016/S0168-9002(99)00498-2}},
\href{http://www.arXiv.org/abs/hep-ex/9902006}{\texttt{arXiv:hep-ex/9902006}}.

\bibitem{ClsCite}
\hrefCMSnoop {}{A.~L. Read, ``Presentation of search results: the ${CL}_s$
  technique'',} \textit{ J. Phys. G} \textbf{ 28} (2002) 2693,
\href{http://dx.doi.org/10.1088/0954-3899/28/10/313}{\doi{10.1088/0954-3899/28/10/313}}.

\end{thebibliography}\endgroup
